\documentclass[12pt,preprint]{aastex}
\usepackage{emulateapj5}

\shorttitle{The Field \& Clusters of NGC 5253}
\shortauthors{Tremonti et al.}

\slugcomment{accepted for publication in ApJ}

\begin{document}

\title{Star Formation in the Field and Clusters of NGC 5253
\footnotemark}
\footnotetext{Based on observations with the NASA/ESA Hubble Space 
Telescope, obtained at the Space Telescope Science Institute, which is 
operated by the Association of Universities for Research in Astronomy, Inc. 
under NASA contract No. NAS5-26555.}

\author{Christy A. Tremonti \altaffilmark{2,3}}
\email{cat@pha.jhu.edu}

\author{Daniela Calzetti\altaffilmark{3}}
\email{calzetti@stsci.edu}

\author{Claus Leitherer\altaffilmark{3}}
\email{leitherer@stsci.edu}
\and

\author{Timothy M. Heckman\altaffilmark{2,3}}
\email{heckman@pha.jhu.edu}

\altaffiltext{2}{Department of Physics \& Astronomy, 
The Johns Hopkins University, 3400 N. Charles Street, Baltimore, MD 21218}
\altaffiltext{3}{Space Telescope Science Institute, 3700 San Martin Drive, 
Baltimore, MD 21218}

\begin{abstract}

We investigate the star formation history of both the bright star clusters
and the diffuse `field star' population in the dwarf starburst galaxy
NGC~5253 using longslit ultraviolet spectroscopy obtained with the Space
Telescope Imaging Spectrograph (STIS). The slit covers a physical area of
370$\times$1.6~pc and includes 8 apparent clusters and several
inter-cluster regions of diffuse light which we take to be the field.  
The diffuse light spectrum lacks the strong O-star wind features which are
clearly visible in spectra of the brightest clusters. This discrepancy
provides compelling evidence that the diffuse light is not reflected light
from nearby clusters, but originates in a UV-bright field star population,
and it raises the issue of whether the star formation process may be
operating differently in the field than in clusters.  We compare our
spectra to STARBURST99 evolutionary synthesis models which incorporate a
new low metallicity ($\sim$1/4~Z$_{\sun}$) atlas of O-star spectra.  The
clusters are well fit by instantaneous burst models with a Salpeter
initial mass function (IMF) extending up to 100~M$_{\sun}$, and we derive
ages for them ranging from 1 to 8~Myrs. Reasonable fits to the field
spectrum are obtained by continuous star formation models with either an
upper mass cut-off of $\sim$30~M$_{\sun}$ or an IMF slope steeper than
Salpeter ($\alpha \sim 3.5$).  We favor a scenario which accounts for the
paucity of O-stars in the field without requiring the field to have a
different IMF than the clusters: stellar clusters form continuously and
then dissolve on $\sim$10~Myr timescales and disperse their remaining
stars into the field. We consider the probable contribution of an O-star
deficient field population to the spatially unresolved spectra of high
redshift galaxies.

\end{abstract}


\section{INTRODUCTION: }

The energy produced in the star formation process is distributed over some
10 decades of frequency, from the X-ray to the radio; however, the bulk of
our knowledge about high redshift (z $\gtrsim$ 2.5)  star forming galaxies
comes from a rather limited observational window: the rest-frame
ultraviolet. This is a consequence of the simple fact that high redshift
galaxies are faint (at z$\sim$3 $L^{*}$ corresponds to R $\simeq$ 24.5,
\citep{Steidel_et_al_1999}) and thus accessible only in spectral regimes
where the sensitivity is optimum.  At present most observations of high-z
galaxies are made by large ground-based telescopes working in the visible,
which corresponds to the rest-frame ultraviolet for galaxies at redshifts
larger than $\sim$2.5.

Fortuitously, the ultraviolet (UV) is an excellent regime in which to
study star forming galaxies:  it contains the direct spectroscopic
signatures of the hot young massive stars which dominate the bolometric
light, as well as an abundance of interstellar features that provide
valuable probes of the gas.  In fact, almost all of the relevant stellar
and interstellar absorption lines have wavelengths below 3000 \AA.  The
full diagnostic power of this rich spectral region is still being
realized: a number of programs are underway to correlate UV spectral
morphology with global galaxy parameters such as mass, metallicity,
extinction, stellar content, and bolometric luminosity
\citep[e.g.][]{Heckman_et_al_1998, Meurer_et_al_1999,
Adelberger_and_Steidel_2000}.  These studies concentrate on local
starburst galaxies whose global properties are well known from
observations in a number of wavebands.

Nearby starburst galaxies are the most obvious local counterparts of the
high redshift galaxies currently being found in large numbers through
color selection based on the Lyman break technique
\citep[e.g.][]{Steidel_et_al_1996, Steidel_et_al_1999}.  While the
extinction corrections to the rest-frame ultraviolet of high-z galaxies
are still a matter of some debate, it is clear that the stellar surface
mass densities and star formation rates per unit area of these galaxies
are several orders of magnitude higher than ordinary late type galaxies,
and are comparable to those of local starbursts \citep{Meurer_et_al_1997,
Kennicutt_et_al_1998}. Local starbursts and Lyman-dropout galaxies share
similar distributions of observed UV colors \citep{Meurer_et_al_1997,
Meurer_et_al_1999, Adelberger_and_Steidel_2000} and have spectral
morphologies which are characterized by blue continua with strong
interstellar absorption, the broad P-Cygni signatures of stellar winds,
and weak Ly-$\alpha$ emission \citep{Steidel_et_al_1996,
Lowenthal_et_al_1997, Pettini_et_al_2000}.

High-resolution UV spectra of local starburst galaxies provide invaluable
templates for comparison with spectra of high-z galaxies.  However,
\emph{such comparisons must be viewed with a note of caution because of
differences in spatial resolution}.  High redshift galaxies have small
angular sizes and spectra which are spatially integrated, whereas local
starbursts have such large angular scales that only select pieces of the
burst can be sampled without using apertures so large as to substantially
degrade the spectral resolution.  Differences in the spatial resolution of
high and low-z observations are particularly significant because bursts of
star formation are known to be highly inhomogeneous. Space-based UV
imaging has shown nearby star forming regions to be composed of compact,
young, UV-bright stellar clusters embedded in a diffuse, irregular, UV
background \citep{Meurer_et_al_1995}.

Historically, high-resolution UV spectroscopy of local starburst galaxies
has focused on individual stellar clusters because of their high surface
brightness.  However, \emph{the UV-bright clusters are not the dominant
contributors to the UV luminosity of a starburst galaxy:} the diffuse UV
light typically accounts for $\sim$ 50\% to 80\% of the total
\citep{Buat_et_al_1994, Meurer_et_al_1995, Maoz_et_al_1996}.  The nature
of the diffuse UV component is hinted at by UV images of the nearest
starbursts where individual high mass stars are resolved
\citep{Meurer_et_al_1995}.  These stars appear to trace the diffuse light
isophotes, suggesting that the diffuse light is not simply reflected light
from the bright clusters, but originates in a UV-bright field star
population.  This raises the issue of whether or not there are two
distinct `modes' of star formation operating in a starburst: one which
produces compact clusters and one which produces more diffuse star
formation.  What is clear is that a starburst is not the same thing as a
star cluster; nor is it the sum of its clusters.  The diffuse light
represents a large, if not dominant, part of the integrated light of a
starburst, and may represent an important and distinct mode of star
formation.

Understanding the nature of the diffuse UV component is one of the
principal aims of our present investigation which utilizes STIS longslit
ultraviolet spectroscopy of the central starburst in the nearby galaxy
NGC~5253.  By addressing the contribution that both the clusters and the
diffuse light make to the integrated spectrum of a local galaxy, we hope
to be able to more accurately interpret the spatially unresolved spectra
of high-z star forming galaxies.

\section{BACKGROUND}

NGC~5253 is a metal-poor dwarf galaxy in the Centaurus Group which
contains an extremely young starburst.  A re-calibration of existing
Cepheid data by the \emph{HST} Key Project on the Extragalactic Distance
Scale team has produced an improved distance estimate to the galaxy of
3.33~$\pm~0.29$~Mpc (\citealt{Gibson_et_al_2000};  
1\arcsec~$\approx$~16~pc).  NGC~5253 has been the target of a number of
observational programs at wavelengths ranging from the X-ray to the radio,
the results of which are summarized by \citet{Caldwell_and_Phillips_1989},
\citet{Martin_and_Kennicutt_1995}, and \citet{Calzetti_et_al_1997,
Calzetti_et_al_1999}.  The metallicity of NGC~5253 is
$\sim\frac{1}{5}$~Z$_{\sun}$ \citep{Kobulnicky_Kennicutt_and_Pizagno_1999}
and is relatively constant across the galaxy, with the exception of a few
areas of enhanced nitrogen abundances \citep{Walsh_and_Roy_1989,
Kobulnicky_et_al_1997}.  NGC 5253's small size and modest metallicity make
it an excellent analogue to galaxies in their early phases of formation,
while its proximity makes it ideal for disentangling the diffuse UV light
from that of the bright stellar clusters.

Our long slit traverses a relatively large portion of NGC 5253, as can be
seen in Figure \ref{image}.  Morphologically the galaxy is somewhat
unusual (termed `amorphous' by \citet{Sandage_and_Brucato_1979}).  
Various authors have suggested that NGC~5253 may have resembled a dwarf
elliptical until an encounter with NGC~5236 (M83) one to two Gyrs ago
triggered the formation of a new stellar population
\citep{Rogstad_et_al_1974, Caldwell_and_Phillips_1989}.  Star formation is
ongoing in the central 20\arcsec\ ($\sim$325 pc) of the galaxy which
contains a dozen UV-bright stellar clusters \citep{Meurer_et_al_1995} and
is very blue except where dust produces patchy, heavy obscuration
\citep{Calzetti_et_al_1997}.  A prominent dust lane with
A$_{V}~\gtrsim~2.2$ mags girds the galaxy nearly perpendicular to its
major axis \citep{Calzetti_et_al_1997}.  Radio observations suggest that a
large fraction of the most recent star formation is hidden by dust
\citep{Turner_Ho_and_Beck_1998}.  The ionized gas in NGC~5253 is more
extended by a factor of $\sim$2 than the continuum emission, and it is
circularly symmetric about a stellar cluster (NGC5253-5 of
\citealt{Calzetti_et_al_1997}) which is located near the geometric center
of the galaxy.

The most active region of star formation is the starburst nucleus, which
is 40 -- 50 pc in size and centered at the position of the H$\alpha$ peak.
An age as young as 5 Myr is suggested by the large H$\alpha$ equivalent
width \citep{Calzetti_et_al_1999}, the presence of Wolf-Rayet stars
\citep{Campbell_et_al_1986, Walsh_and_Roy_1987, Walsh_and_Roy_1989,
Schaerer_et_al_1997}, the small number of red supergiants
\citep{Campbell_and_Terlevich_1984}, and the purely thermal component of
the radio emission \citep{Beck_et_al_1996}.  Stars in this nuclear region
are still embedded in the parental cloud with A$_{V} \sim$~9 -- 35 mag
\citep{Calzetti_et_al_1997}.  Recent observations at 1.3 and 2 cm have
revealed the presence of a $\sim$1--2~pc `supernebula' --- a very young
and gas-rich counterpart of a super-star cluster --- which may be the
youngest globular cluster known \citep{Turner_Beck_and_Ho_2000}.  The star
formation rate density in the nuclear region is between $10^{-5}$ and
$10^{-4}~$M$_{\sun}$~yr$^{-1}$~pc$^{-2}$ \citep{Calzetti_et_al_1999},
which corresponds to the maximum levels observed in starburst galaxies
\citep{Meurer_et_al_1997}.  However, the bulk of the observed UV light is
not produced by the starburst nucleus, but by the surrounding regions
(extending out to 250 pc), which are older, relatively unextincted, and
producing stars at a rate 10 times lower than that of the nuclear
starburst \citep{Calzetti_et_al_1997}.

\section{OBSERVATIONS AND DATA REDUCTION} \label{observations}

We obtained longslit ultraviolet spectra of the central region of NGC~5253
with the Space Telescope Imaging Spectrograph (STIS) aboard the
\emph{Hubble Space Telescope} on 1999 July~27.  We utilized the 52\arcsec\
$\times$~0.1\arcsec\ slit and the G140L grating to obtain the best
compromise between throughput and spectral resolution.  The spectra span a
wavelength range of 1150 -- 1700 \AA\ with a resolution of $\sim$2~\AA,
which provides a good match to our stellar evolutionary synthesis models
(1200 -- 1600~\AA;  $\delta\lambda \sim 1.5$~\AA).

\begin{figure}
\plotone{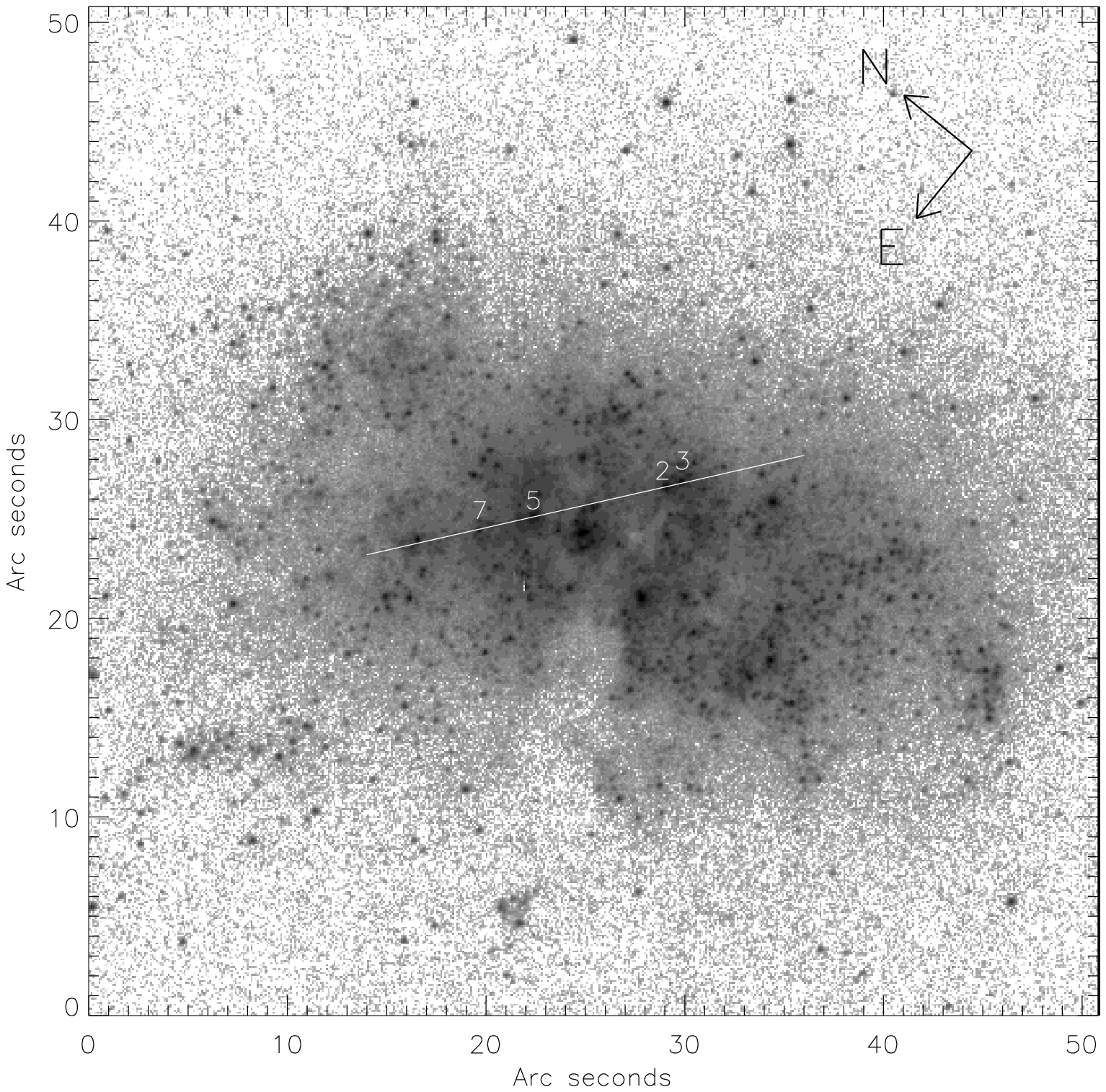}
\caption{A \emph{Hubble Space Telescope} WFPC2 image of
NGC~5253 in the F255W filter.  The approximate placement of the slit is
shown, and Clusters 2, 3, 5, and 7 are labeled for reference. One
arcsecond corresponds to $\sim$16 pc.
\label{image}}
\end{figure}

The position of the slit was chosen so as to intersect Clusters~2, 3, and
5 of Calzetti et al. 1997 (see Figure \ref{image}).  Target acquisition
was accomplished in a two phase process utilizing the STIS CCD at visible
wavelengths.  First, a bright foreground star 1\farcm4 from Cluster~5 was
acquired in a 5\arcsec\ field of view and centered on the CCD to an
accuracy of $\pm$ 0\farcs01. An offset was then performed to Cluster~5
(RA~=~13\fh39\fm56\fs02, Dec~=~-31\fdg38\arcmin25\farcs0, J2000)  and a
peakup was done to refine the centering.  The spacecraft moved the slit
across the target parallel to the dispersion direction in five 0\farcs075
steps.  The optimal spacecraft location (accurate to 5\% of the slit,
$\sim$0\farcs005) was determined from a flux weighted centroid algorithm.
The slit orientation (PA~=~56\fdg246) was dictated by Clusters~2 and 3,
which are separated by $< 1$\arcsec\ and are $\sim$8\arcsec\ distant from
Cluster~5.

Following the successful target acquisition, the G140L grating was moved
into the beam and four frames of duration 1410, 2945, 2945, and 2945
seconds were obtained with the STIS FUV-MAMA detector.  The data were
processed with the Space Telescope Science Institute's \emph{CALSTIS}
pipeline, which provides automated rebinning, global detector linearity
correction, dark subtraction, flat fielding, wavelength calibration, and
conversion to absolute flux units.  For our subsequent analysis we
utilized the fully calibrated two-dimensional images produced by the
pipeline, which are rectified to be linear in both the wavelength and
spatial dimension. After examining the four calibrated frames for spatial
shifts, we combined them into a single two-dimensional image using the
IRAF task mscombine.

\begin{figure} 
\plotone{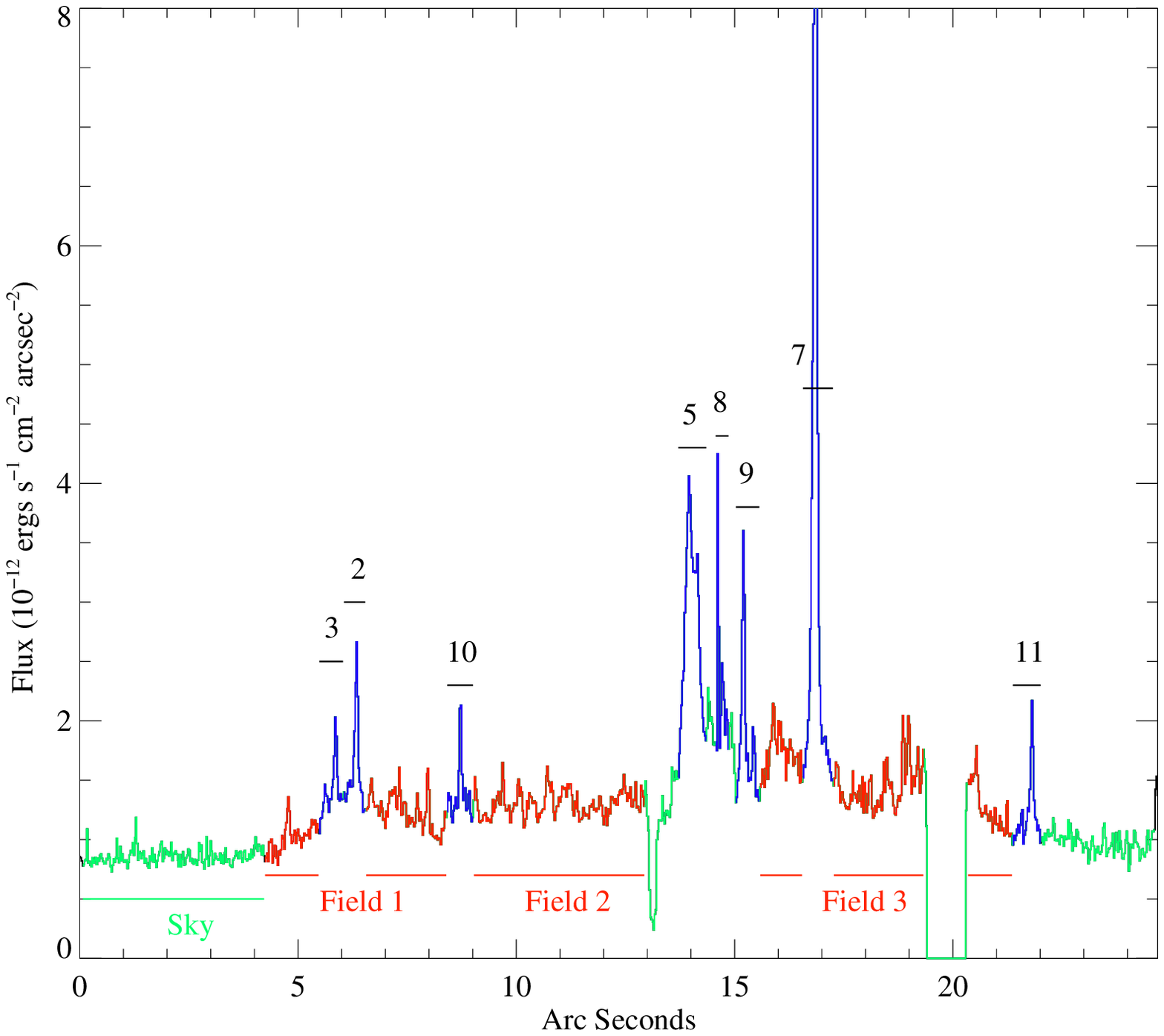} 
\caption{The integrated 1150 -- 1710 \AA\ flux as a function
of position along the slit.  Note the plateau of diffuse light extending
from roughly 4 -- 22\arcsec\ along the slit. The regions defined as
clusters are shown in blue and labeled 2 -- 11, in accordance with the
convention of Calzetti et al. (1997).  The three regions which we take to
be the field are shown in dark red. The portion of the spectrum used for
the subtraction of the geocoronal lines is labeled ``sky''. Horizontal
bars indicate the extent of the extraction windows. The areas where the
spectrum falls to zero are artifacts of the STIS occulting bar and
repeller wire.
\label{spatial}}  
\end{figure}

The slit traverses roughly 370~pc of NGC~5253, including both the intense
regions of star formation near the nucleus and the more quiescent outer
regions (see Figure \ref{image}). The spatial dimension of the data
extends 23\arcsec\ with an image scale of 0\farcs0244~pixel$^{-1}$
($\sim$0.4~pc~pixel$^{-1}$), which is sufficient to resolve stellar
clusters.  Figure~\ref{spatial} shows the integrated 1150 -- 1710~\AA\
flux as a function of slit position.  A number of bright point-like
sources are visible, embedded within a $\sim$300~pc plateau of diffuse
light. For our subsequent analysis, we define the 8 brightest peaks along
the spatial axis to be clusters, and take regions between those peaks to
be the field or diffuse component.  (We use the terms `cluster' and
`field' quite liberally here.)  Data within these subregions were binned
into one-dimensional spectra.

\begin{figure}
\plotone{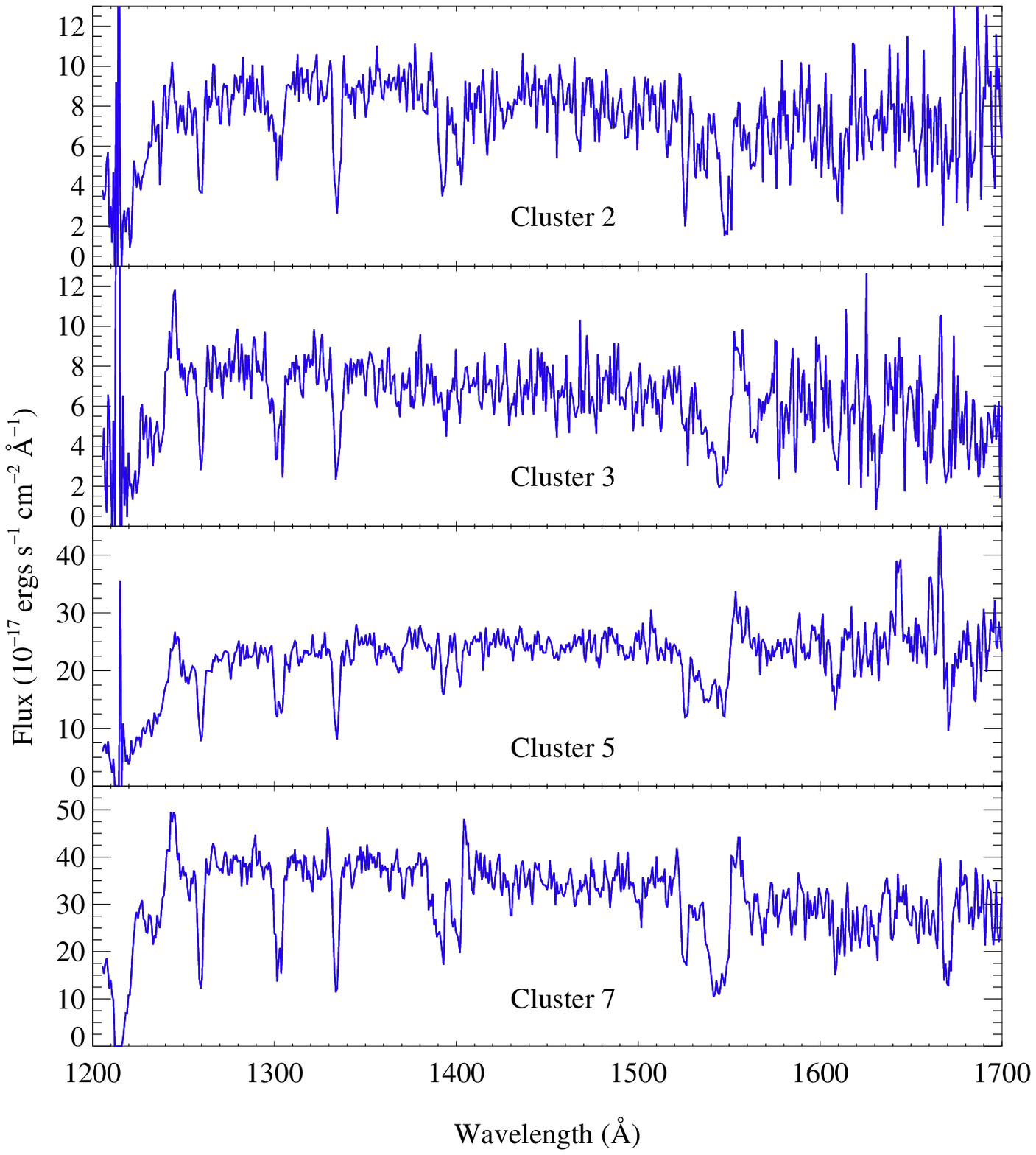}
\caption{Spectra of the four brightest clusters 2, 3, 5, and
7 are shown for comparison.  These data have not been corrected for
reddening or slit losses.
\label{bright_clusters}}
\end{figure}

The spatial pixels binned up to form a given cluster's spectrum were
chosen by eye from an examination of the spectrally integrated data shown
in Figure~\ref{spatial}.  The spectral properties of the fainter clusters
were fairly sensitive to the boundaries chosen.  We made a concerted
effort to balance the competing demands of signal-to-noise (S/N) and
spectral purity by examining the effects that broadening the extraction
had on the resultant spectrum.  Some of our so-called clusters may
nevertheless include more than one coeval population (note the two
component blended profile of Cluster~5 in Figure~\ref{spatial}).  The
extracted spectra of the four brightest clusters are shown in
Figure~\ref{bright_clusters}.

\begin{figure}
\plotone{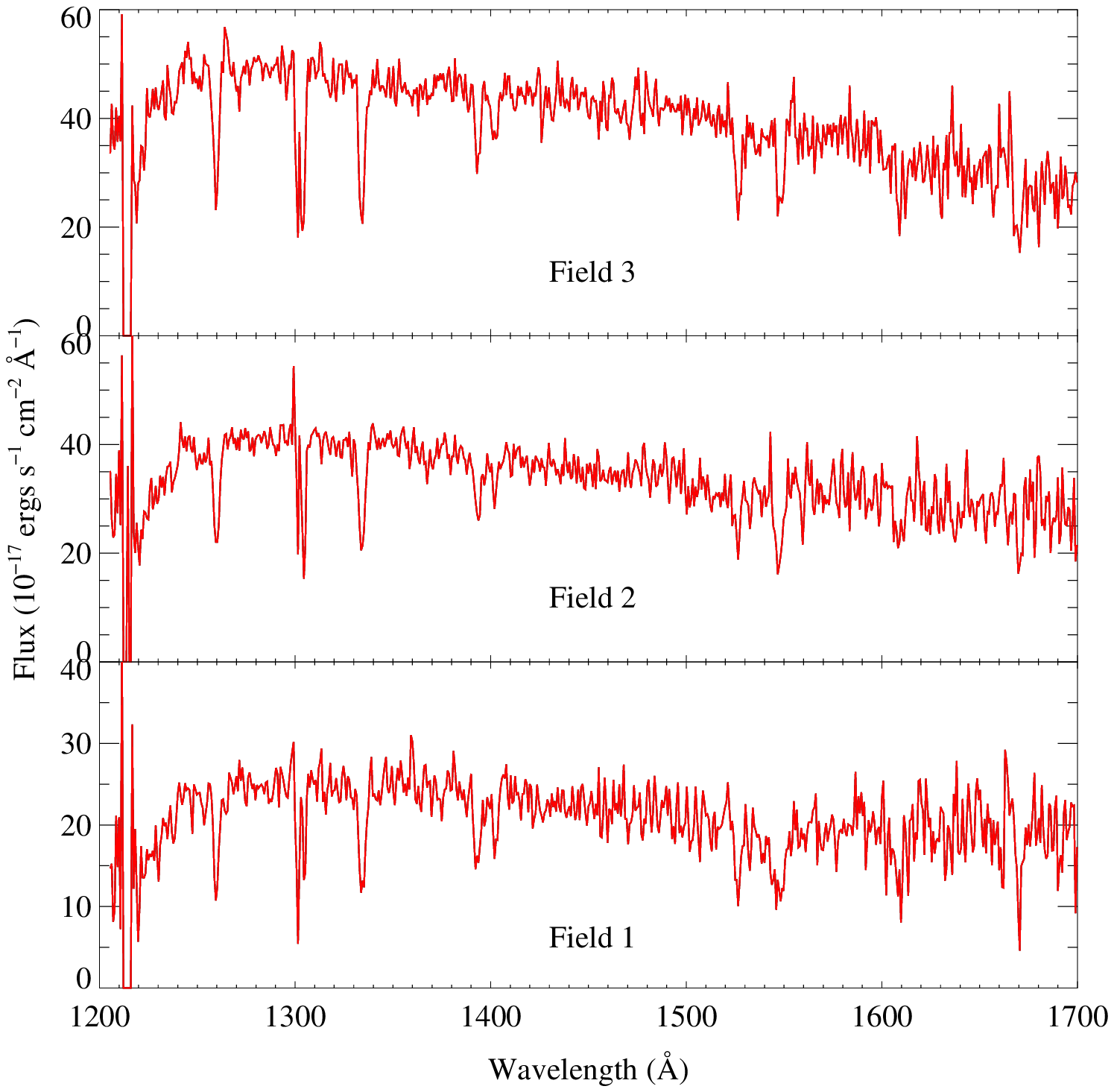}
\caption{Spectra of the three field regions are shown for 
comparison.  The data have not been corrected for reddening. 
Imperfect subtraction of the geocoronal  Ly-$\alpha$ and \ion{O}{1} 
$\lambda$1302 lines produces the residual spikes seen near Ly-$\alpha$
and the spurious emission feature near 1300~\AA\ in Field~2.
\label{fields}}
\end{figure}

Because of their greater extent, the field regions were not nearly as
sensitive to their spatial boundaries.  However, subtle differences were
apparent when large areas were considered. We therefore defined three
distinct field regions (shown in Figure~\ref{spatial}), each with a S/N
comparable to that of the brightest clusters.  The extracted field spectra
are shown in Figure \ref{fields}.

The low redshift of NGC~5253 ($v= 404$~km~s$^{-1}$) precluded direct
measurement of our spectral resolution because the strong interstellar
features of the starburst are blended with intervening Milky Way
absorption lines.  We therefore estimate our resolution based on the known
properties of STIS.  At the distance of NGC~5253, a typical cluster is
neither a point source nor a fully extended source, so we expect our
spectral resolution to be intermediate between point and extended source
values (0.9~\AA\ and 2.4~\AA\ respectively).  We estimate a value of
$\sim$1.8~\AA\ as an appropriate resolution for our cluster spectra.  Our
field spectra should have the resolution of a fully extended source.  We
measure a 2.2~\AA\ FWHM for the Lyman-$\alpha$ airglow line.

Because the clusters are embedded in a diffuse background and are
resolved, we treated all of our spectra as extended continuum sources when
converting from units of surface brightness in the two-dimensional image
to units of flux (ergs~s$^{-1}$~cm$^{-2}$~\AA$^{-1}$) in the
one-dimensional extracted spectra.  Treating the clusters as point sources
would increase their measured fluxes by approximately a factor of two due
to corrections for the extended wings of the PSF.

The extracted spectra were corrected for slight relative wavelength shifts
(0.02 -- 0.2 \AA) based on the peak geocoronal Ly-$\alpha$ emission,
de-redshifted, and then rebinned to the model resolution of 0.75~\AA\ per
pixel. We then corrected the spectra for the strong geocoronal emission at
Ly-$\alpha$ and \ion{O}{1}~$\lambda1302$.  Data in a region at the edge of
the diffuse light plateau (the region labeled ``sky'' in Figure
\ref{spatial}) were extracted to produce a one-dimensional spectrum.  
While this spectrum included substantially more pixels than any of the
field regions, the stellar features were extremely faint, making it a
suitable ``sky'' spectrum.  We subtracted a weak continuum from the sky
spectrum and set all the pixels to zero except for a 40~\AA\ window around
Ly-$\alpha$ and a 20~\AA\ window around \ion{O}{1}. Taking this to be a
purely geocoronal spectrum, we then subtracted it from the 1-dimensional
spectra of the cluster and field regions in proportion to the number of
pixels covered by each.  The intrinsic interstellar absorption lines of
\ion{O}{1} and \ion{Si}{2} near $\lambda 1302$~\AA\ appear to be fairly
well recovered.

Extinction corrections were applied individually to each of the 
unrectified spectra.  We removed the foreground Galactic extinction with 
the standard Milky Way extinction law of 
\citet{Cardelli_Clayton_and_Mathis_1989}, using a reddening of E(\bv) = 0.05
\citep{Burstein_and_Heiles_1982} for all of the spectra.  
The extinction intrinsic to the starburst was removed using the starburst
obscuration curve \citep[and references therein]{Calzetti_et_al_2000}.  
Assuming a foreground dust geometry with a standard extinction curve (e.g.
the Milky Way or LMC curve) would generally give a similar or lower color
excess than the values derived with the starburst obscuration curve.  
However, the differences would be small in the present case, owing to the
modest reddening affecting the stellar clusters (with the exception of
Cluster~5 which is discussed further in \S \ref{clusters}, and possibly
Cluster~8.)

We used $\beta$, the power law slope of the UV continuum ($F \propto
\lambda^{\beta}$), as a diagnostic of the reddening internal to the
starburst.  Because of the plethora of weak lines in the spectra, the
value of $\beta$ measured was fairly sensitive to the method of continuum
fitting.  We used 10 iterations of a sigma rejection routine with a
2.0$\sigma$ lower bound and a 3.5$\sigma$ upper bound to perform a power
law fit over the wavelength range 1240 -- 1600 \AA.  This method produces
systematically shallower slopes than the method of
\citet{Calzetti_et_al_1994} in which a fit is performed over 10 line-free
windows (only 5 of which are in our wavelength range.)  However, in this
instance we are not concerned with the absolute value of $\beta$, merely
the difference in $\beta$ between the data and the models, which are free
of reddening.  For a wide range of model parameters (detailed in \S
\ref{modeling}), we measure $\beta$ values of -2.50 to -2.75.  Since our
measurements are probably not accurate to more than $\pm 0.1$, we take
-2.6 to be the intrinsic slope of the UV continuum for all of our spectra.
We derive E(\bv) values for each of the cluster and field regions by
de-reddening the observed spectrum using the starburst obscuration law
until the measured value of $\beta$ is -2.6.  The observed $\beta$ values
for each of the cluster and field regions are listed in
Tables~\ref{table_cluster_data} and \ref{table_field_data}.
Table~\ref{table_cluster_results} contains the $\beta$ values of the
clusters corrected for galactic foreground reddening and the derived color
excess.

\section{MODELING}\label{modeling}

We compare the spectra of the clusters and the field to the
\emph{STARBURST99} stellar evolutionary synthesis models
\citep{Leitherer_et_al_1999}.  The \emph{STARBURST99} models (hereafter
SB99) have been optimized to reproduce many spectrophotometric properties
of galaxies with active star formation.  We briefly describe those aspects
of the models which we utilize in \S \ref{SB99}; in \S \ref{modelpars} we
describe our input model parameters; and in \S \ref{modelfitting} we
describe our fitting routine.

\subsection{STARBURST99} \label{SB99}

The SB99 models use two limiting cases of the star-formation law: an
instantaneous burst with no subsequent star formation and star formation
proceeding continuously at a constant rate.  Stellar populations are
parameterized by their stellar initial mass function (IMF), which is
assumed to be a power law
($\Phi~\propto~\int_{M_{low}}^{M_{up}}m^{-\alpha}dm$)  between a lower
mass cut-off (M$_{low}$) and an upper mass cut-off (M$_{up}$).  The
stellar population is evolved from the ZAMS using the evolutionary models
of the Geneva group.  Luminosity, effective temperature, and radius are
calculated for each star during each evolutionary time step
(10$^{4}$~years).  Spectral types are determined, and an appropriate
ultraviolet spectrum (1200 -- 1600~\AA) is assigned to each star from a
stellar library.  The resultant model spectra are simply a combination of
the constituent stellar spectra at each time step.

The library spectra utilized by SB99 are composite stellar spectra with
signal-to-noise ratios of $\sim$20. The original library consisted of
spectra created from high dispersion (0.75 \AA) IUE observations of solar
or slightly sub-solar metallicity Milky Way stars, with O and Wolf-Rayet
star spectra drawn from the atlas of
\citet{Robert_Leitherer_and_Heckman_1993} and B-star spectra from
\citet{de_Mello_et_al_2000}.  In this study we take advantage of 
an important and very recent addition to the
SB99 library: an atlas of metal poor ($\sim$1/4 Z$_{\sun}$) O-star
spectra created from HST observations of stars in the Large (LMC) and
Small (SMC) Magellanic Clouds \citep{Leitherer_et_al_2001}.

\subsection{Model Parameters} \label{modelpars}

We compare our data to SB99 models with a variety of star formation
scenarios.  The high mass stellar content of a starburst depends upon both
the slope of the initial mass function, $\alpha$, and the upper mass
cut-off of the IMF, M$_{up}$.  Populations which form in an instantaneous
burst will also have a dependence on age, since the most massive stars are
the shortest lived. We construct three grids of models where two of these
three parameters ($\alpha$, M$_{up}$, and age) are held fixed and the
third is allowed to vary.

Our benchmark model is an instantaneous burst with a Salpeter IMF slope
$\alpha$~=~2.35, a lower mass cut-off of M$_{low}$~=~1~M$_{\sun}$, and an
upper mass cut-off of M$_{up}$~=~100~M$_{\sun}$.  We allow this model to
evolve from 0 -- 50 Myrs and create a grid of spectra with ages in 1 Myr
increments.  The parameters of our instantaneous burst model are in line
with results from recent studies that have shown the high-mass end
(M$\ge$5~M$_{\sun}$) of the IMF to be remarkably uniform in both resolved
regions of active star formation \citep{Massey_1998, Massey_et_al_1995b,
Hunter_et_al_1996} and in starbursts \citep{Garcia-Vargas_et_al_1995,
Stasinska_and_Leitherer_1996}.  The high-mass end slope in stellar
clusters is measured to be $\alpha\approx$~2.3--2.5
\citep[e.g.,][]{Scalo_1998}, which is, for all practical purposes, a
`Salpeter slope', and the upper mass cut-off is $>$70--100~M$_{\sun}$
\citep{Massey_1998}.  No systematic variations in the IMF slope are
observed over about a factor 10 in metallicity and a factor $\sim$1000 in
stellar density \citep{Massey_et_al_1995a, Massey_et_al_1995b}. In this
study we are not particularly concerned with the low end of the IMF,
except for cluster mass determinations (see \S~\ref{clusters}),
as our UV spectroscopy is insensitive to the low-mass cut-off. We have
tested that low-end-truncated IMFs up to 10~M$_{\sun}$ do not induce a
significant variation in the UV spectral features of our models, and,
thus, in the cluster age determinations.

We also consider continuous star formation models.  Continuous models take
approximately 10 Myr to equilibrate, after which the UV spectrum changes
very little as a function of age.  We therefore fix the age at 50~Myr and
let the free parameters be the IMF slope and the upper mass cut-off. We
create 2 grids of continuous models: one with $\alpha = 2.35$,
M$_{low}$~=~1, and M$_{up}$~=~10,~20,~30,~40,~50,~100~M$_{\sun}$ and the
other with M$_{low}$~=~1~M$_{\sun}$, M$_{up}$~=~100~M$_{\sun}$, and
$\alpha$~=~2.35,~3,~3.5,~4,~5.  Continuous star formation models should be
more appropriate to star formation occurring in spatially extended regions
such as the stellar field, where the crossing times are long, typically of
order 10~Myr.

For all of our SB99 models, we used an evolutionary track metallicity of
0.2~Z$_{\sun}$ (Z$_{\sun}$~=~0.02, by mass) and the new LMC/SMC stellar
libraries. This provides a good match to the metallicity of NGC~5253's gas
which is known to be $\sim\frac{1}{5}$~Z$_{\sun}$.  The good agreement
between the metallicity of our data and that of the the stellar libraries
is particularly important because the age and IMF sensitive features also
have a Z-dependence, with the general trend being towards weaker lines at
lower metallicity \citep{Leitherer_et_al_2001}.  In particular, the
strength of the P-Cygni features depends upon the mass loss rate which is
a function of metallicity \citep[e.g.][]{Puls_et_al_1996}. This dependence
is evident in comparisons of stars of the same spectral type and
luminosity class in the Milky Way, LMC, and SMC
\citep[e.g.][]{Garmany_and_Conti_1985, Walborn_et_al_1995,
Walborn_et_al_2000}.

One important caveat is that the low metallicity library employed by SB99
extends only from O3 through B0, with all other spectral groups being drawn
from the original solar metallicity library.  This metallicity mismatch
between the high mass stars and intermediate and low mass stars can affect
the computed synthetic spectrum of a composite population depending on its
age or IMF.  The line profiles of the stellar wind lines, which are
produced primarily in very massive stars ($>40$~M$_{\sun}$), are not much
affected, while the continuum and many photospheric lines are more
sensitive to contributions from B stars.  \citet{Leitherer_et_al_2001}
conclude that for populations with a standard Salpeter IMF up to 100
M$_{\sun}$ the effect of the solar metallicity B-stars is relatively minor
for continuous star formation models or instantaneous bursts with ages
less than $\sim10$ Myr. This is discussed further in \S \ref{field}.

\subsection{Model Fitting} \label{modelfitting}

\begin{figure} 
\plotone{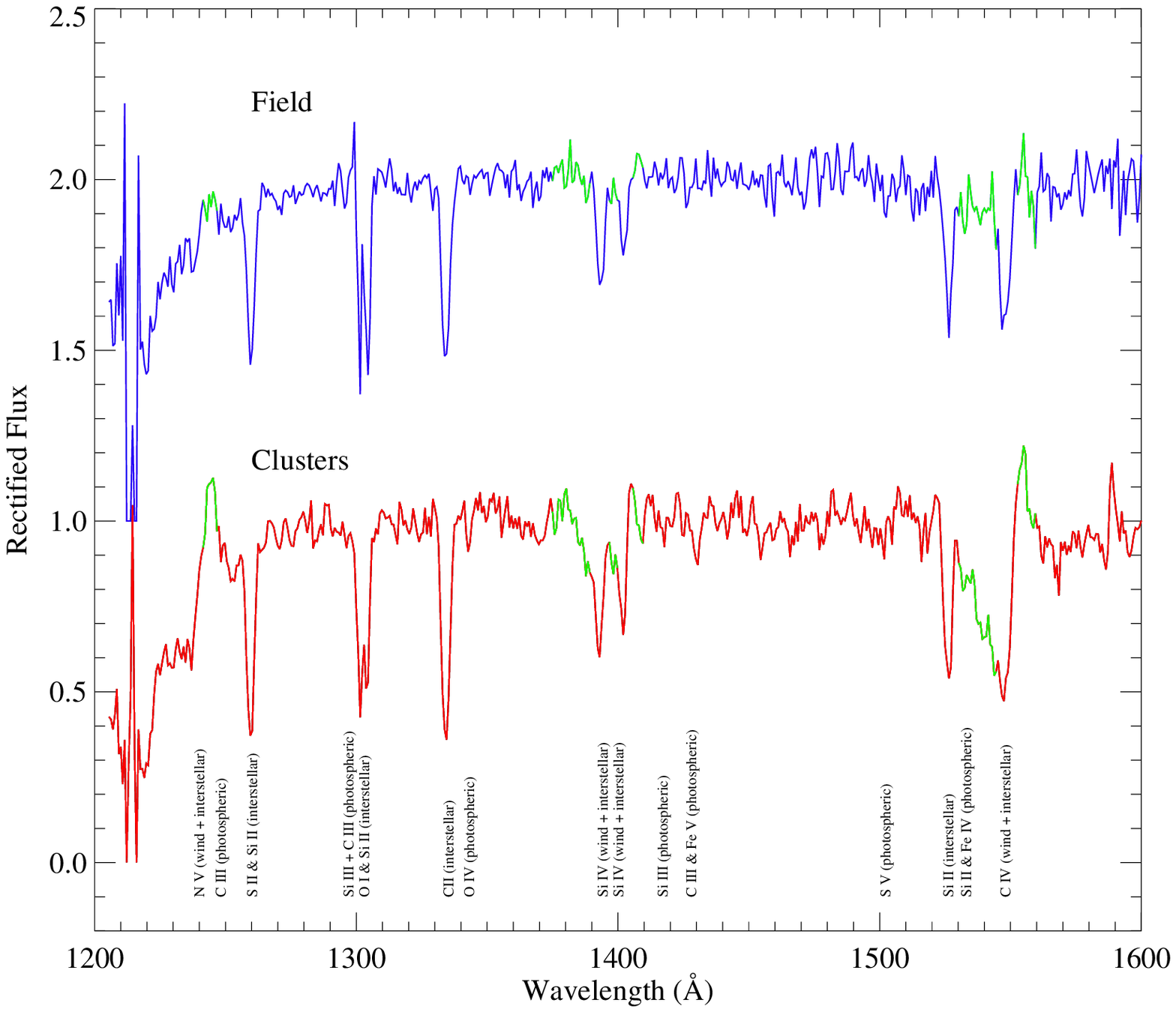} 
\caption{The rectified spectrum of the field (top) is shown
with the combined spectrum of the clusters (bottom) for comparison.  The
cluster spectrum shows the clear signatures of O-stars in the the broad
stellar wind P-Cygni features of \ion{N}{5} $\lambda1240$, \ion{Si}{4}
$\lambda1400$, and \ion{C}{4} $\lambda1550$ (highlighted in green.)  The
corresponding features are greatly reduced in the B-star dominated field
spectrum.  A number of stellar wind and photospheric features are labeled
as are the prominent interstellar lines, which are a blend of the
intrinsic starburst lines and foreground Milky Way absorption.  
We note that the spikes seen near Ly-$\alpha$ and the emission 
feature seen near 1300~\AA\ in the field spectrum are residuals from the 
geocoronal line subtraction.
The regions highlighted in grey are the stellar wind lines with their
interstellar cores removed. These regions are especially sensitive to the
age/stellar composition of the population and are heavily weighted in
determining the best fit of evolutionary synthesis model spectra to the
data.  
\label{field_cluster}} 
\end{figure}

The ultraviolet spectrum of a star forming region provides a sensitive
measure of its high-mass star content because the most massive stars are
also the most luminous, and they dominate the integrated light.  Figure
\ref{field_cluster} contrasts the O-star dominated composite cluster
spectrum with the B-star dominated spectrum of the field.  An O-star
dominated spectrum is characterized by the broad ($\sim$2000~km~s$^{-1}$)
stellar wind P-Cygni profiles of \ion{N}{5}~$\lambda1240$,
\ion{Si}{4}~$\lambda1400$, and \ion{C}{4}~$\lambda1550$.  These features
are notably absent in B-star dominated spectra, which exhibit increasingly
strong narrow ($\sim$350~km~s$^{-1}$) photospheric features, such as
\ion{Si}{2}~$\lambda1265$ and \ion{Si}{2}~$\lambda1309$.  
Our data have sufficient S/N ($\sim$10) and spectral
resolution ($\sim$2~\AA) to allow us to utilize these stellar features as
probes of the massive star content in various regions of NGC~5253.

We determined which models provided the best fit to our cluster and field
spectra in the 1200 -- 1600~\AA\ regime using an automated routine that
takes as input the extinction corrected spectrum of a cluster or field
region, a grid of SB99 model spectra, and an array of values corresponding
to the free variable in the grid of models (i.e. the age, IMF slope, or
M$_{up}$.)  For each of the model spectra in the grid the goodness of the
fit is characterized by a $\chi^{2}$ value computed in the following
manner:  $\chi^{2} = (o_{i} - m_{i})^{2} w_{i} / \sigma_{i}^{2}$, where
$o_{i}$ represents the observed data for the $i$th pixel, $m_{i}$ the
model data, $\sigma_{i}$ the error in the observed spectrum, and $w_{i}$
the assigned weight.

Weights were applied in order to maximize the sensitivity of the
$\chi^{2}$ to regions of the spectrum which are most influenced by the
age/stellar content of the burst, namely the stellar wind lines. On the
basis of the detailed line analysis of de Mello et al. (2000), we
classified each pixel in our spectra as belonging to either the continuum,
an interstellar line, or a stellar wind line, and assigned weights
accordingly.  The interstellar lines were given a weight of zero; the 470
pixels of continuum a weight of 1; and the 69 pixels corresponding to
stellar wind lines a weight of 10.  The result of this weighting scheme
was to eliminate the interstellar lines from consideration, and to give
the wind lines a weight $\sim1.5$ times greater than that of the
continuum.

It was necessary to eliminate the interstellar absorption lines from our
$\chi^{2}$ because SB99 is designed to model the integrated UV light of a
purely \emph{stellar} population; no consideration is given to the gaseous
environment of the star-forming region.  Interstellar absorption lines are
present in the models only in so much as they are present in the library
of stellar spectra.  Unfortunately, nearly all of the features which are
sensitive diagnostics of the stellar population have some interstellar
contamination. In the case of the broad stellar wind features, we attempt
to mask out the narrow interstellar core (see Figure~\ref{field_cluster}).  
However, the separation of stellar and interstellar features remains a
source of uncertainty.

\begin{figure}
\plotone{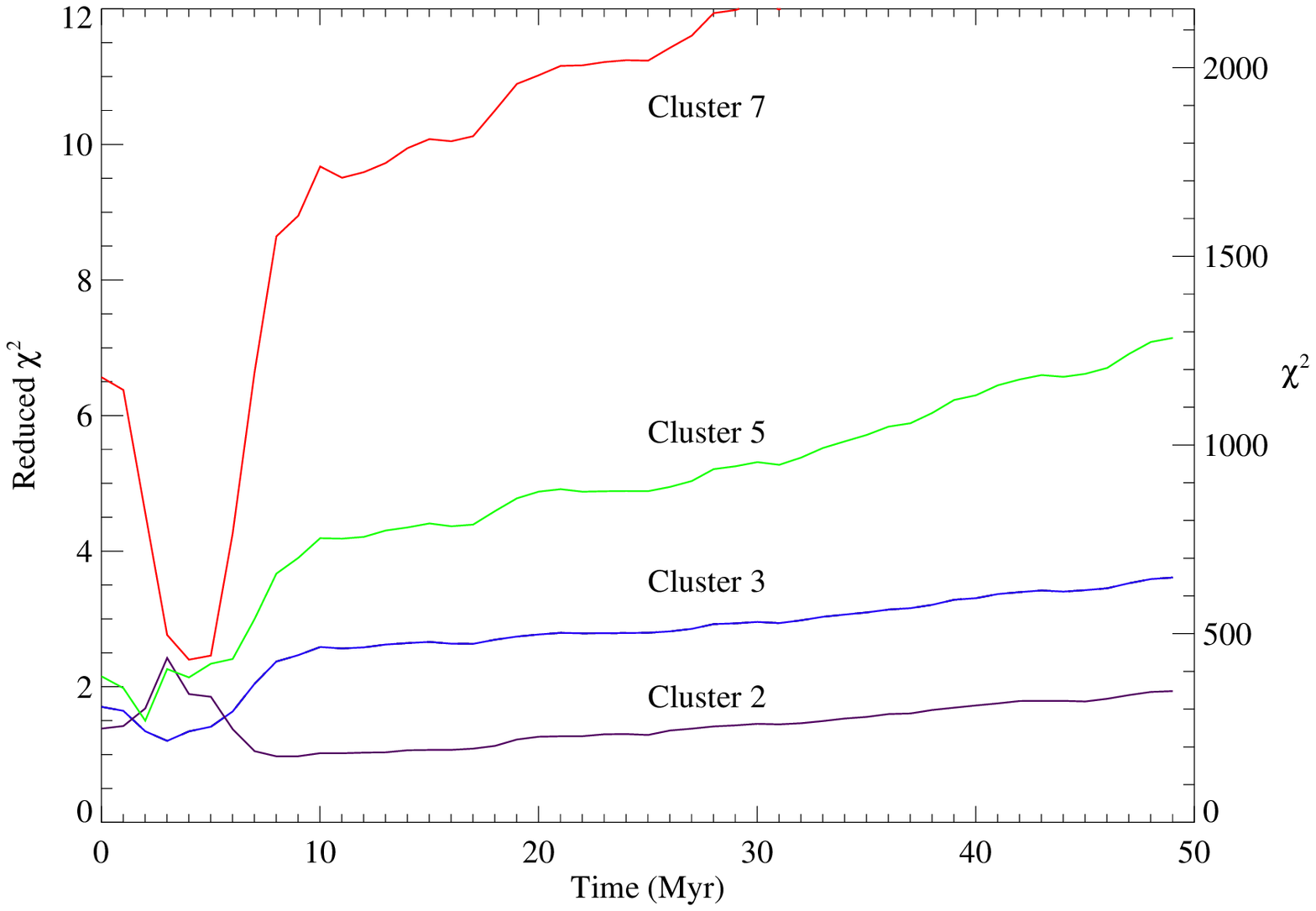}
\caption{The $\chi^{2}$ evolution of an instantaneous burst
model fit to Clusters 2, 3, 5, and 7 is shown.  The shape of the
$\chi^{2}$ provides information about the robustness of the fit.  
Cluster~5, for example, has an age which is tightly constrained to be near
2~Myr by its steep $\chi^{2}$ slope on either side of minimum. In
contrast, the evolution of the $\chi^{2}$ for Cluster 2 is relatively flat
after $\sim8$~Myr, indicating a large uncertainty on the upper bound of
its age.
\label{chisq}}
\end{figure}

Our routine returns the best fit model (minimum $\chi^{2}$) and plots the
evolution of the $\chi^{2}$ with the free parameter, as shown in Figure
\ref{chisq}. The best fits proved to be relatively robust to the method of
$\chi^{2}$ evaluation: small changes in the weighting scheme employed had
virtually no effect, especially in cases where the S/N was good or O-star
wind lines were still present.  In order to quantify the error associated
with our best fit models, we utilized the bootstrap method. This entailed
randomly resampling the residuals of the best fit, adding these to the
model spectrum, and running the resultant spectrum through our automated
fitting routine.  This procedure was repeated 1000 times for each
spectrum.  The error bars associated with the 90\% confidence interval
were derived from a histogram of the results.

\section{ANALYSIS}

\subsection{Clusters} \label{clusters}

The regions of the spectrum that we define as clusters are shown in Figure
\ref{spatial}.  Our definition of what constitutes a `cluster' is clearly
somewhat arbitrary: we have taken the eight brightest peaks in the
spectrum.  Our clusters are not necessarily bound entities, but we retain
this labeling for simplicity.  For consistency we use the cluster naming
convention of \citealt{Calzetti_et_al_1997} for Clusters~2, 3, and 5, and
number the remaining clusters (starting with 7) in order of decreasing
observed flux.  The extracted spectra of the four brightest clusters are
shown in Figure~\ref{bright_clusters}.  The measured properties of the
clusters are reported in Table~\ref{table_cluster_data}.

All of our clusters, except for Cluster~8, have spatial profiles that are
clearly resolved (their guassian FWHM values are well in excess of the 1.6
pixels expected for a point source at 1400~\AA) and relatively symmetric.
Cluster~8 is actually a collection of a few peaks of UV emission covering
13~pixels (5.2~pc).  The most prominent of these peaks extends no more
than four pixels and is compatible with a point source.  Its UV
luminosity, L$_{1500} \simeq 5 \times 10^{35}$~ergs~s$^{-1}$, could be
produced by a single O-supergiant, but the observed spectral features are
more consistent with those of a B-star.  We calculate that 3--4 B
supergiants could produce the observed UV luminosity, implying that
Cluster~8 is actually an association with a modest number of stars.
  
\begin{figure}
\plotone{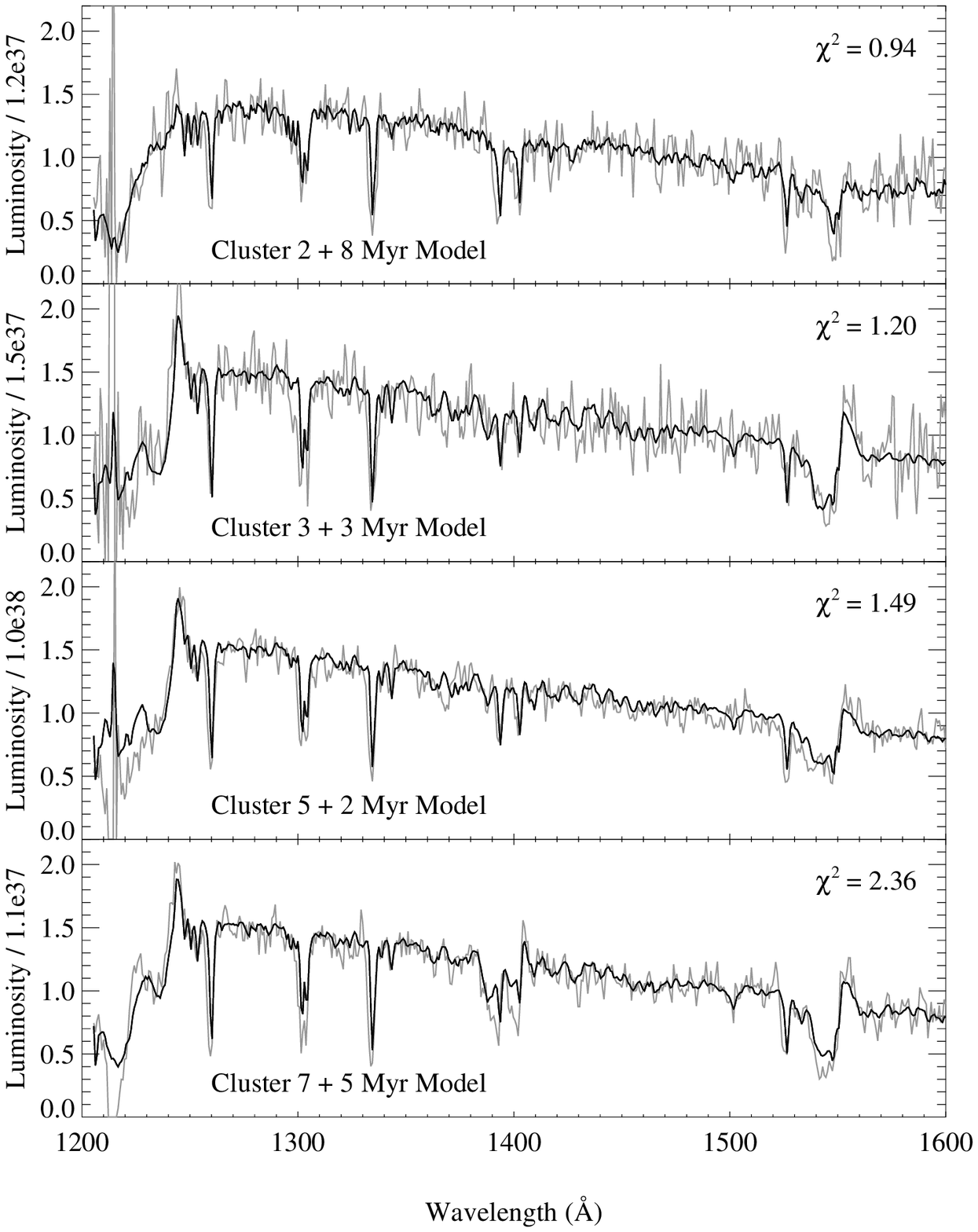}
\caption{Dereddened spectra of the four brightest clusters 2,
3, 5, and 7 are shown in grey with their best fitting instantaneous burst
models overplotted in black.  Luminosity units are
ergs~s$^{-1}$~\AA$^{-1}$.  The reduced $\chi^{2}$ values of the fit are
labeled on the plots for reference.
\label{clustermodels}}
\end{figure}

We compared our dereddened cluster spectra to a grid of instantaneous
burst models with a Salpeter IMF slope ($\alpha = 2.35$), an upper mass
cut-off of 100~M$_{\sun}$, and ages ranging from 0 to 50 Myr in 1 Myr
increments.  An instantaneous burst model is appropriate for the clusters
since they have typical half light radii of 0.6~parsecs, and thus have
negligibly short crossing times, of order 0.1~Myr.  All eight cluster
spectra were remarkably well fit by the models. The four brightest
clusters are shown with their best fitting models in
Figure~\ref{clustermodels}. We assign ages to the clusters on the basis of
their best fits and error bars on these ages which represent the 90\%
confidence interval (see \S \ref{modelfitting} for details).  The
robustness of our age determination can also be assessed by examining the
shape of the $\chi^{2}$ evolution, shown in Figure~\ref{chisq} for
Clusters 2, 3, 5, and 7.  The ages of the other clusters are expected to
be more uncertain than the formal error bars reflect since they are less
luminous and potentially subject to strong stochastic effects if only a
handful of O-stars are present. The ages and other model dependent
properties of the clusters are summarized in
Table~\ref{table_cluster_results}.

We derive cluster masses from the dereddened luminosity of the clusters.
For the 4 brightest clusters (2, 3, 5, 7) we derive fluxes from photometry
of an archival WFPC2 image in the F170W filter. The signal-to-noise of the
image (a combination of two 400 second images) is too low for the
remaining clusters to be clearly detected.  We have corrected our F170W
image for warm pixels, decontamination events, geometric distortion, and
charge transfer efficiency.  The photometry we obtain provides a more
accurate estimate of the total luminosity of the clusters since they are
extended objects and their exact position relative to the slit is unknown.
In the most extreme case, Cluster~3, the luminosity derived from the
photometry is a factor of $\sim30$ greater than that obtained from our
STIS spectrum. The masses of the four clusters too faint to be detected in
the F170W image (8, 9, 10, 11) are therefore lower limits.

All of the clusters have been de-reddened as described in \S
\ref{observations} using the starburst obscuration curve
\citep{Calzetti_et_al_2000}.  This assumption provides a lower limit to
the actual dust obscuration (and hence a lower limit to the dereddened UV
luminosity and mass) if more complex geometries, such as dust mixed with
stars, are present. To clarify the magnitude of the effect we use as
an example the most reddened cluster in our sample, Cluster~5. Cluster~5
is located in the heart of what \citet{Calzetti_et_al_1997} refer to as
the ``starburst nucleus'' of NGC~5253.  Its measured UV slope is $\beta =
0.06$ (after correction for the small amount of reddening from our own
Galaxy), much redder than the expected slope of a 2~Myr old cluster,
$\beta\simeq -2.71$. Using the starburst obscuration curve, the color
excess implied by the difference between the slopes is E(\bv) = 0.42.  
The resulting intrinsic UV luminosity is L$_{1500} = 1.0 \times
10^{38}$~ergs~s$^{-1}$~\AA$^{-1}$, taking the flux measured from the F170W
image (to avoid slit losses) and assuming a distance of 3.33~Mpc.  We use
the number of ionizing photons reported by SB99 for a 2 Myr old cluster
scaled by L$_{1500}^{obs}$/L$_{1500}^{model}$ and the transformations of
\citet{Leitherer_and_Heckman_1995} to predict an extinction corrected
H$\alpha$ luminosity of L$_{H\alpha}\simeq3.6\times10^{39}$~erg~s$^{-1}$.  
This value is in reasonable agreement with the measured value of
\citet{Calzetti_et_al_1997}, which is
L$_{H\alpha}\simeq1.3\times10^{40}$~erg~s$^{-1}$ after applying the same
obscuration correction used above and taking into account the fact that
the nebular gas is more reddened than the stellar flux by a factor of
$\sim2.3$ \citep{Calzetti_et_al_2000}).

Unfortunately, this ``dust scenario'' is degenerate with a number of
others, in the sense that multiple dust geometries can reproduce the
observed quantities (UV slope and H$\alpha$ flux), the only difference
being the prediction of larger intrinsic luminosities, and hence the
cluster masses.  For instance, if we embed Cluster~5 in a dust cloud with
A$_V$=35~mag \citep{Calzetti_et_al_1997} with A$_V\simeq$0.2 of foreground
dust, we recover the observed UV slope and H$\alpha$ luminosity while
predicting a cluster mass 10 times greater than that implied by the
previous scenario. The dust cloud geometry can not be constrained by
diagnostics at UV-optical wavelengths; for example, the maximum value of
the color excess that is measurable from the Balmer decrement
(H$\alpha$/H$\beta$) in these cases is E(B$-$V)$\lesssim$0.35
\citep{Calzetti_1997}. Only by going to much longer wavelengths, such as
the radio, can the degeneracy be broken \citep{Beck_et_al_1996,
Calzetti_1997}.

Cluster~5 has recently been mapped with the NRAO Very Large Array in the A
configuration at wavelengths of 1.3 and 2~cm
\citep{Turner_Beck_and_Ho_2000}.  The authors claim no optical counterpart
for their source; however, the position of the dominant radio source is
only 1.17\arcsec\ distant from our pre-peak up coordinates for Cluster~5,
which is within the error bars for absolute position measurements from
\emph{HST}. The radio source is partially optically thick at 2~cm,
implying a Lyman continuum rate of 3 $\times 10^{52}$ s$^{-1}$
\citep[corrected to 3.33~Mpc] {Turner_Beck_and_Ho_2000}, which is about 3
times higher than that predicted on the basis of the observed dereddened
H$\alpha$ flux.  We estimate a mass of 4~$\times10^5$~M$_{\sun}$ for
Cluster 5 by comparing the measured Lyman continuum rate with that
reported by SB99 for a 2 Myr old 10$^6$~M$_{\sun}$ cluster with a Salpeter
IMF extending from 1--100~M$_{\sun}$. This value is a factor of $\sim$10
above the mass we derive solely on the basis of our UV measurements.  
However, it should be kept in mind that Cluster~5 is an extreme case: we
expect luminosity and mass uncertainties to be substantially lower for
most of the other clusters in our sample, since they are considerably less
reddened.

Another factor which can contribute to underestimated masses for the
stellar clusters is the adopted lower mass limit to the IMF. Direct
measurements \citep{Sirianni_et_al_2000, Luhman_et_al_2000} and
morphological and dynamical information \citep{Meurer_et_al_1995,
Ho_and_Filippenko_1996} favor a significant low-mass ($<$1~M$_{\sun}$)
population in stellar clusters.  However, due to the observed flattening
of the IMF at low masses, the total mass in stars should be similar to
that derived using a Salpeter IMF extending down to 1~M$_{\sun}$.  If we
take a more extreme assumption, for instance, if the IMF follows a
Salpeter slope down to $\sim$0.3~M$_{\sun}$, the clusters will be a factor
$\sim$1.6 more massive than what is reported in Table
\ref{table_cluster_results}.

\subsection{Field} \label{field}

We investigated the diffuse light of NGC~5253 by creating spectra of three
`field' regions.  We choose these regions such that they excluded any
apparent clusters, but fell within the plateau of diffuse light (see
Figure~\ref{spatial}).  The resultant one-dimensional field spectra have
signal-to-noise values of 12 -- 18, comparable the brightest
clusters.  The three field spectra are shown in Figure~\ref{fields} and their 
measured properties are reported in Table~\ref{table_field_data}. 
Since the three regions have similar spectral morphologies and amounts of 
reddening, we combine them into a 
single field spectrum which we use for all of our subsequent analysis.

The field spectrum does not show any evidence of the broad line profiles
of \ion{N}{5}, \ion{Si}{4}, and \ion{C}{4} which are characteristic of
O-star winds.  This is in sharp contrast to the spectra of the brightest
clusters.  The obvious disparity between the cluster and field spectra
(see Figure \ref{field_cluster}) is clear evidence that the diffuse light
is not simply reflected light from the clusters, but must originate in a
UV bright field star population.

We devote our subsequent analysis to the nature of the field star
population.  What is immediately apparent from the weakness of the O-star
wind features is that the field is deficient in the most massive stars.  
We consider whether this could be a statistical effect due to the small
area and modest star formation rate of the region we are sampling.  If
we assume that the light in the field is the result of a continuous star
formation episode, SB99 predicts that $\sim10$ O-stars should be present
for a Salpeter IMF extending from 1 to 100~M$_{\sun}$.  (We have assumed a
burst duration of 50~Myr, although the result is largely insensitive to
this value.)  The salient point is that a non-negligible number of O-stars
is expected to be present. The fact that we do not see the spectral
signatures of these high mass stars in our field spectrum indicates that
either the IMF slope is steeper than Salpeter, the upper mass cut-off is
lower than 100~M$_{\sun}$, or that the stellar population has aged past
$\sim$10~Myr.  We create models to explore each of these scenarios.

\begin{figure}
\plotone{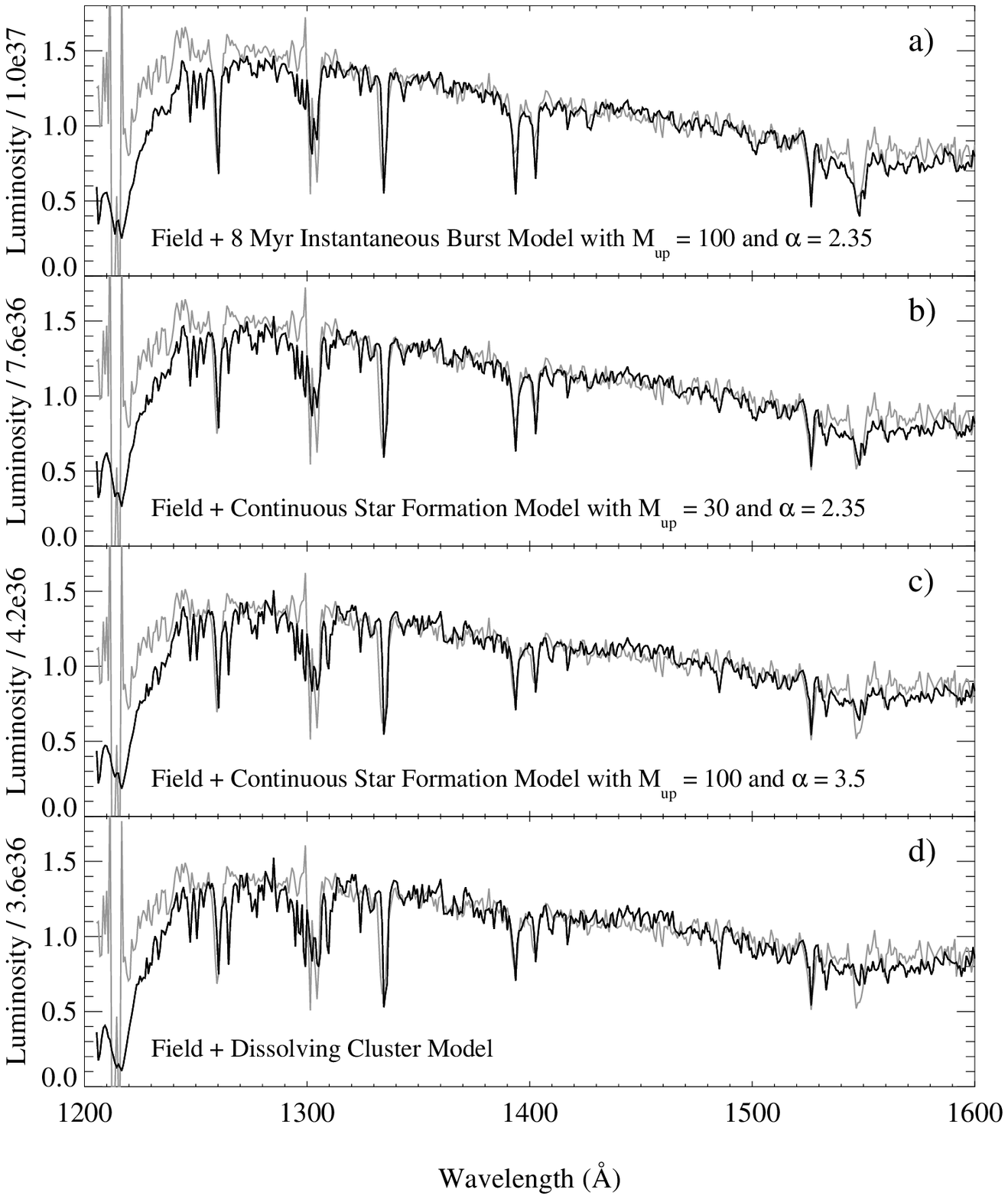}
\caption{The total field spectrum is shown in grey with
different models overplotted in black: a) an 8 Myr old instantaneous burst
model with a Salpeter IMF extending up to 100 M$_{\sun}$ b) a continuous
star formation model with a Salpeter IMF slope ($\alpha = 2.35$) and an
upper mass cut-off of M$_{up} = 30$~M$_{\sun}$ c) a continuous star
formation model with $\alpha = 3.5$ and M$_{up} = 100$~M$_{\sun}$ d) 100
Myrs of continuous star formation with 10 Myrs of continuous star
formation subtracted off.  This model represents a ``dissolving clusters''
scenario (see \S~\ref{fielddiscussion}.) Luminosity units are
ergs~s$^{-1}$~\AA$^{-1}$.
\label{fieldmodels}}
\end{figure}

We compare our various field models to the data using the same automated
routines that we used for the clusters.  However, we do not
quote formal error bars on our best fits because the field spectrum has
such high S/N that systematic errors dominate, as is evident in
Figure~\ref{fieldmodels}~.  The probable sources of systematic error are
discussed subsequently.  

The power law slopes of models which are very deficient in massive stars are 
much shallower than our assumed value of -2.6 used for the 
clusters. In order not to bias our fits against these models, 
we rectify both the data and the models by fitting and
dividing by the appropriate power law before comparison.  
Shallower intrinsic values of $\beta$ imply less reddening in the data.  This
provides a useful constraint because models can not be redder that the
extreme case of zero reddening for the field, which corresponds to $\beta
= -1.56$, after correction for foreground Galactic reddening.  

The first model we consider is the fiducial model that we use for the
clusters: an instantaneous burst with a Salpeter IMF slope
($\alpha$~=~2.35) and an upper mass cut-off of M$_{up}$~=~100~M$_{\sun}$.
This model produces a best fit to the composite field spectrum at an age
of 8 Myr.  Younger ages are strongly ruled out by the increasing strength
of O-star wind lines.  At ages older than about 8 Myr, the primary age
discriminator becomes the strength of the B-star photospheric lines.  
However, the strength of these lines is expected to increase with
metallicity, so the use of the solar metallicity B-star library could bias
our fit towards younger ages.  Thus, 8 Myr should be viewed as a lower
bound to the age of the field.  A strict upper bound of 45 Myr is set by
the reddest model slope allowable, corresponding to the rather unlikely
case of E(\bv)~=~0. Ages of 8 - 18 Myr imply an E(\bv) value of $\sim0.17$,
in good agreement with the reddening of most of the clusters flanking the
field regions. However, it is not implausible that the field could be less
reddened than the stellar clusters.

The second model we consider is a continuous star formation model with a
fixed age of 50~Myr, a Salpeter IMF slope, and variable upper mass
cut-off. We test values of M$_{up}$ of 10, 20, 30, 40, 50, and 100
M$_{\sun}$, and find an upper mass cut-off of 30~M$_{\sun}$ is preferred.
Higher upper mass cut-offs are strongly ruled out by the increasing
presence of the O-star wind lines, particularly \ion{C}{4}.  Values as low
as 10~M$_{\sun}$ are unlikely because of the extreme strength of the B star
photospheric lines (which are barely evident in the field spectrum).  
However, such models 
can not be conclusively eliminated because of the unknown impact of the solar
metallicity B-star library on these features.

The third model we consider has continuous star formation with the age
fixed at 50~Myr, the upper mass cut-off fixed at 100~M$_{\sun}$, and a
variable IMF slope.  We examine values of $\alpha$ = 2.35, 3, 3.5, 4, and
5. The steepest slope we consider, $\alpha = 5$, corresponds to the `field
star IMF' derived by \citet{Massey_et_al_1995a} for the LMC and SMC. This
model is ruled out a priori by the fact that it has an intrinsic slope of
$\beta = -0.89$, far redder than the value measured for the field after
correction for foreground reddening ($\beta = -1.56)$. The best fitting
model, $\alpha = 3.5$, has a slope of $\beta = -2.1$, implying a very
modest reddening of E(\bv) = 0.08.  IMF slopes shallower than this are
ruled out by the increasing width of the \ion{C}{4} stellar wind line.

The field spectrum is shown with each of the three model spectra
overplotted in Figure~\ref{fieldmodels}~(a--c).  None of the models
provides a satisfactory fit. In each case, the continuum level of the data
is above that of the models at wavelengths shortward of 1270 \AA\ and
longward of 1530 \AA. In addition, photospheric lines like [\ion{S}{3}]
$\lambda\lambda$1294,1296,1298~\AA\ are grossly overpredicted.  One
consideration is that our subtraction of the geocoronal emission at
Ly-$\alpha$ and [\ion{O}{1}] $\lambda1302$ could be inducing errors in
these regions of the spectrum.  A residual emission feature is clearly 
observed in the field spectrum near 1300~\AA.  However, other 
photospheric features such as \ion{Si}{2} $\lambda$1260, which are 
uncontaminated by geocoronal lines, are poorly matched with the models as 
well. It is also worth noting that the cluster data seem 
to be reasonably well fit in the regions surrounding the geocoronal lines, 
suggesting that sky subtraction is not a major source of error.  

Another possibility is that the discrepancies between the data and 
the models are the result of a metallicity mismatch, as the stars with masses 
less than 20 M$_{\sun}$ (the predominant contributors to the continuum and
photospheric lines in these models)  have empirical spectra drawn from the
solar metallicity library. At solar metallicity, the broad wing of
Lyman-$\alpha$ is more extended, and increased line blanketing is expected
near 1550 \AA, which could account for discrepancies in the continuum
level. Also, photospheric lines are observed to be substantially weaker at
LMC/SMC metallicity than at solar metallicity
\citep{Leitherer_et_al_2001}, consistent with what is seen in the field.  
However, until low metallicity B-stars spectra are added to the SB99
library, we can not be sure if this fully accounts for the discrepancy.  
An intriguing possibility is that the field spectrum might contain
contributions from an older underlying stellar population.  Adding a burst
with an age of $\sim1$~Gyr to the models can eliminate the continuum
mismatch problems at both ends; however, the already strong photospheric
features become even stronger, and an unrealistically large stellar mass
is implied. Until the metallicity issues are resolved, conclusions about
an underlying older population are premature.

The parameters which provided the best fit for each of the three models
are listed in Table \ref{table_fieldmodels} along with the measured power
law slope of the model, and the implied reddening. The reduced $\chi^{2}$
values of the best fits are also given for reference.  We do not use these
values to discriminate among the models, since the differences are
relatively minor and the overall quality of the fits is not high.  We
consider the plausibility of each of the models further in \S
\ref{fielddiscussion}.

\section{DISCUSSION}\label{discussion}

\subsection{The Stellar Clusters: Ages and Masses}\label{clusterdiscussion}

The eight clusters subtended by our slit all have ages below $\sim$10~Myr,
with three having ages $<$3~Myr, for a Salpeter IMF in the range
1--100~M$_{\sun}$. Among the three youngest is Cluster~5, the primary
target of our STIS observations, whose very young age is in agreement with
that inferred by \citet{Calzetti_et_al_1997} based on the large observed
equivalent width of the H$\alpha$ emission line (EW(H$\alpha$)=1650~\AA).

Notably young are Clusters~2 and 3, with ages 8~Myr and 3~Myr,
respectively. For these two clusters, age estimates based on colors give
values of $\sim$60~Myr and $\sim$30~Myr, respectively
\citep{Calzetti_et_al_1997}, much larger than what is inferred from UV
spectroscopy. Age determinations based on colors alone are indeed bound to
be relatively uncertain unless additional constraints (e.g., H$\alpha$
emission) are available.  For instance, consider Cluster 2:  the
difference in \bv color between a 60 and a 8 Myr old instantaneous burst
is only $\sim0.18$~mag. The narrowness of this range makes accurate
reddening corrections imperative.  \citet{Calzetti_et_al_1997} deredden
their observed colors based on the ratio of H$\alpha$ to H$\beta$; however
the measured equivalent widths are very small ($<4$~\AA) and therefore
bound to be uncertain. After correction for galactic foreground reddening
and underlying stellar absorption, they find zero magnitudes of foreground
attenuation between H$\alpha$ and H$\beta$.  By contrast, the reddening
derived from our UV spectroscopy is E(\bv)~=~0.16, very nearly what is
needed to bring the broad-band color age estimate into line.  Similar
results are found for Cluster 3, although the \bv color difference between
the two age estimates is greater ($\sim$0.3) and can not be entirely
reconciled by applying our reddening correction. Additional corroboration
for the young ages of Clusters 2 and 3 comes from an analysis of the soft
X-ray emission near the center of NGC~5253
\citep{Strickland_and_Stevens_1999}. The two clusters appear to be located
within a superbubble of expanding hot gas. Energy limitations constrain
the age of the superbubble to be $\lesssim$10~Myr
\citep{Strickland_and_Stevens_1999}, in better agreement with the age
estimates from UV spectroscopy.

The measured and derived properties of the clusters are summarized in
Tables \ref{table_cluster_data} and \ref{table_cluster_results}.  The ages
of the clusters show no particular spatial correlation (which could
indicate propagating star formation.)  Their half-light radii span a wide
range of values (0.6~--~2.4~pc), but visual examination of the profiles
suggests that the largest clusters are an unresolved blend of smaller
ones.  The clusters display varying amounts of reddening, ranging from
E(\bv)~=~0.42~--~0.05.  The reddening values show a broad spatial
correlation; that is, clusters which are physically close show similar
amounts of reddening.  The extinction-corrected 1500~\AA\ luminosities are
in the range L$_{1500}\sim10^{35}$--$10^{38}$~ergs~s$^{-1}$~\AA$^{-1}$ for
the eight clusters, implying masses in the 10$^2$--10$^4$~M$_{\sun}$
range, for a low-end IMF mass of 1~M$_{\sun}$. However, both luminosities
and masses listed can generally be considered lower limits to the actual
values, as discussed in \S \ref{clusters}.  We note that the lowest mass
clusters ($10^{3}$~M$_{\sun}$ or less) might be subject to strong
stochastic effects since only a handful of O-stars are expected to be
present for a Salpeter IMF.  These effects could potentially bias our age
and mass determinations for Clusters~9, 10, and 11.

\subsection{The Stellar Field: The Final Stop for the Evolution of Clusters?}
\label{fielddiscussion}

The UV emission from the field regions surrounding the clusters has a
clearly stellar origin and is not due to reflected light from the clusters
themselves, since, as already mentioned in \S\ref{field}, the spectral
features of clusters and field are notably different. In contrast to the
clusters spectrum, the field spectrum lacks the strong O-star wind lines
of \ion{N}{5}, \ion{Si}{4}, and \ion{C}{4} which are signatures of the
most massive stars. We stress that this deficiency is not an artifact of
undersampling of the field, as our field spectrum contains a considerable
amount of light. In fact, the field has a flux nearly equivalent to the
integrated cluster spectrum, which contains over $6 \times
10^{4}$~M$_{\sun}$ of stars.  
A similar discrepancy between the cluster
and field population has been observed in the starburst galaxy NGC~1569.  
Using WFPC2 photometry, \citet{Greggio_et_al_1998} find that the stellar
population of NGC~1569 is composed of recently formed super star clusters
and resolved field stars with ages greater than 10~Myr.

In order to understand why the field might be deficient in massive stars
we have compared the field spectrum of NGC~5253 to three different
families of model spectra.  The first is an instantaneous burst model with
a Salpeter IMF in the 1--100~M$_{\sun}$ range.  Our best fit model has an
age of 8 Myr, although this should be considered a lower limit for the
reasons discussed in \S\ref{field}.  While this model provides a
reasonable fit to the data, we rule it out on physical grounds: our field
regions are spatially extended ($\sim80$~pc) and therefore unlikely to be
forming stars in an instantaneous burst.  The subsequent models we
consider have continuous star formation which is probably more appropriate
given the galaxy's $\sim10$~Myr crossing time.

Since the age of the burst is largely irrelevant in continuous models, we
allow the free parameters to be the slope and upper mass cut-off of the
IMF.  Our second model uses a fixed Salpeter slope and a variable upper
mass cut-off.  A reasonable fit is produced for an upper mass cut-off 
of $\sim$30~M$_{\sun}$.  The third model fixes the upper mass
cut-off at 100 and allows the IMF slope to vary.  The best fit is achieved
by a value of $\alpha = 3.5$, which is steeper than Salpeter, but
shallower than the $\alpha\sim$5 value found by \citet{Massey_et_al_1995a}
for the stellar field of the Magellanic clouds.  An IMF slope of $\alpha =
5$ is strongly ruled out by the observed blue color of the field's UV
continuum (although it might be argued that our definition of the `field'
is not as rigorous as Massey's.) We note that our fits are sensitive to
stars with masses $\gtrsim$20~M$_{\sun}$ which is comparable to the
stellar mass range $\sim$15--60~M$_{\sun}$ sampled by
\citet{Massey_et_al_1995a}.

The basic result is that the field of NGC~5253 is generally lacking very
massive stars, either because of aging or because very massive stars do
not form in large numbers.  \citet{Meurer_et_al_1995} found that the field
accounts for a significant fraction of the UV light from starbursts,
between 50\% and 80\%, and proposed that the field is the outcome of
diffuse star formation, a mode different from the one which produces
stellar clusters. If the hypothesis that two modes of star formation exist
in starbursts is correct, then our observations show that the two modes
produce drastically different IMFs: a `regular' Salpeter IMF up to
$\sim$100~M$_{\sun}$ in clusters, and a massive-star-deficient IMF in the
field.

A scenario other than bi-modal star formation can be suggested to explain
the properties of the stellar field. The eight clusters along our STIS
slit are all younger than $\sim$10~Myr; the other three bright clusters in
the center of NGC~5253 are younger than $\sim$20~Myr, although for two of
the clusters ages have been derived from photometry alone and can
therefore be fairly uncertain \citep{Kobulnicky_et_al_1997,
Schaerer_et_al_1997, Calzetti_et_al_1997}.  If the clusters along our slit
are representative of the general cluster population in the center of
NGC~5253, we can conclude that the field appears generally older than the
clusters or as old as the oldest among them. The field thus contains stars
at a more advanced evolutionary stage than the clusters. One way to obtain
such a population mix is to hypothesize that stellar clusters dissolve
over timescales of $\sim$10--20~Myr and their surviving stars disperse
into the field.  Thus, there is only one star formation mode:  stars form
in clusters and a fraction of them continue their life in the field.

To test this hypothesis, we have constructed a model for the field by
subtracting a 10~Myr constant star formation population from a 100~Myr
constant star formation population. Star formation has been going on for a
while in the center of NGC~5253 \citep[e.g.,][]{Calzetti_et_al_1997}, and
if clusters dissolve over a timescale of $\sim$10~Myr, the field will
consist of stars with ages between $\sim$10~Myr and the maximum age of the
star formation episode, which we take to be 100 Myr, although the
resulting UV spectrum is largely insensitive to this value. Figure
\ref{fieldmodels} d) compares the field to the $100-10$~Myr model
spectrum. The similarity between this model spectrum and the other models
shown in Figure \ref{fieldmodels} is not surprising: the UV spectrum of
the $100-10$~Myr model is dominated by its youngest, most luminous
contributors, i.e., stars with ages $\sim$10~Myr which have just come out
of dissolving stellar clusters.

One prediction of the $100-10$~Myr model is that about 40\% of the UV
light at 1500~\AA~ is in the field. This fraction becomes slightly higher
as one goes to longer wavelengths, for instance, at 2200 \AA\ (the
wavelength of the \citet{Meurer_et_al_1995} study) the fraction becomes
45\%.  However, this value is still somewhat below the 50--80\% value
derived for nearby starbursts \citep{Buat_et_al_1994, Meurer_et_al_1995,
Maoz_et_al_1996}.  Pushing the dissolution timescale to an age as young as
6 Myr, brings the fraction of light in the field to $>$50\%.  However, the
differences between the measured and predicted fraction of light in the
field may simply reflect the observational difficulty of disentangling
small clusters from the field population.

Star clusters are known to dynamically evolve, with a number of mechanisms
contributing to stellar depletion eventual cluster dissolution. However, the
dissolution timescales implied by our model for NGC~5253 are very short.  
We consider the feasibility of these timescales using the model results of
\citet*{Kim_Morris_and_Lee_1999}. \citet*{Kim_Morris_and_Lee_1999}
calculated the evaporation timescales for compact (half-mass radius
$\le$1~pc) stellar clusters near the center of the Milky Way, including
stellar evolution among the ingredients. With the possible exception of 5,
our clusters appear fairly compact, with half-light radii $<$2~pc; we
assume that the half-light radii are a reasonable approximation of the
half-mass radii, although mass segregation may have already acted on the
oldest clusters to make the first radius smaller than the second. The
environment surrounding our clusters is also quite different from that of
the Milky Way. NGC~5253 has about 100 times less baryonic mass than the
Galaxy, spread over about 20 times less volume (adopting flattened mass
distributions for both galaxies, with similar scale heights); this implies
that the central density in NGC~5253 is about 5 times lower than the
central density in the Milky Way.  \citet{Kim_Morris_and_Lee_1999}
identify the tidal radius as one of the crucial parameters which
determines the evaporation timescale of the stellar clusters; in NGC~5253
the stellar clusters have tidal radii about 70\% bigger than in the Milky
Way, for constant cluster mass, and, thus, evaporation timescales about
2.2 times longer. A 5$\times$10$^3$~M$_{\sun}$ cluster placed 100~pc away
from the dynamical center of NGC~5253 will evaporate in $\sim$15~Myr for a
1--150~M$_{\sun}$ Salpeter IMF or in $\sim$25~Myr for a
0.1--150~M$_{\sun}$ Salpeter IMF. We are therefore in the expected
ballpark timescale for the less massive among our clusters to agree with
the dissolving clusters scenario; heavier clusters, like 2, will have
lifetimes in the range of 50--100~Myr and will not contribute
significantly to the UV emission of the field. The most massive of our
clusters, Cluster~5, has a relaxation time of 400~Myr for a
1--150~M$_{\sun}$ Salpeter IMF, and is potentially bound if the IMF
extends to lower stellar masses.

A final consideration is the validity of our definitions of `cluster' and
`field'.  The first issue of concern is that some of the less massive
objects which we have defined as clusters may simply be random
superpositions of a few field O-stars.  In this scenario our finding of an
massive star deficient field population would be merely circular logic
since we would have selectively eliminated many of the O-stars from the
field.  The brightness of Clusters 9, 10, and 11, while lower limits,
could be consistent with being one to a few O-stars or B supergiants.  
However, all of these clusters are well resolved and have relatively
smooth spatial profiles.  (Cluster 8, which may include a bright point
source, was previously discussed in \S~\ref{clusters}.)  An additional
check is performed by adding the light from Clusters 8, 9, 10, and 11 to
our field spectrum. These four clusters contribute about 20\% of the total
field light, but produce a negligible difference in the integrated
spectrum.  Thus, we can be relatively secure that our choice of clusters
is not robbing the field of O-stars.

The other issue of concern is the converse:  that what we call the field
may in fact include a number of faint unresolved clusters.  Taking the
flux of our faintest cluster as our detection limit, we find that a
10$^4$~M$_{\sun}$ cluster with an E(\bv) of 0.2 would be detectable with
an age as old as 50 Myr, while a 10$^3$~M$_{\sun}$ cluster with a similar
amount of reddening would only be seen at ages less than 15~Myr.  
Clusters with masses less than 10$^3$M$_{\sun}$ probably do not contribute
much UV light since they are unlikely to contain many very massive stars
due to statistical effects. Thus, clusters which are both relatively old
and not very massive could contribute somewhat to our field population.  
However, such light-weight clusters would be subject to a number of
destruction mechanisms, and would tend to dissolve on timescales shorter
than those discussed above \citep{Kim_Morris_and_Lee_1999}.  Most
importantly, the addition of faded clusters to the field would not alter
the basic notion that there is only one `mode' of star formation.

\subsection{High Redshift Galaxies}

\begin{figure}
\plotone{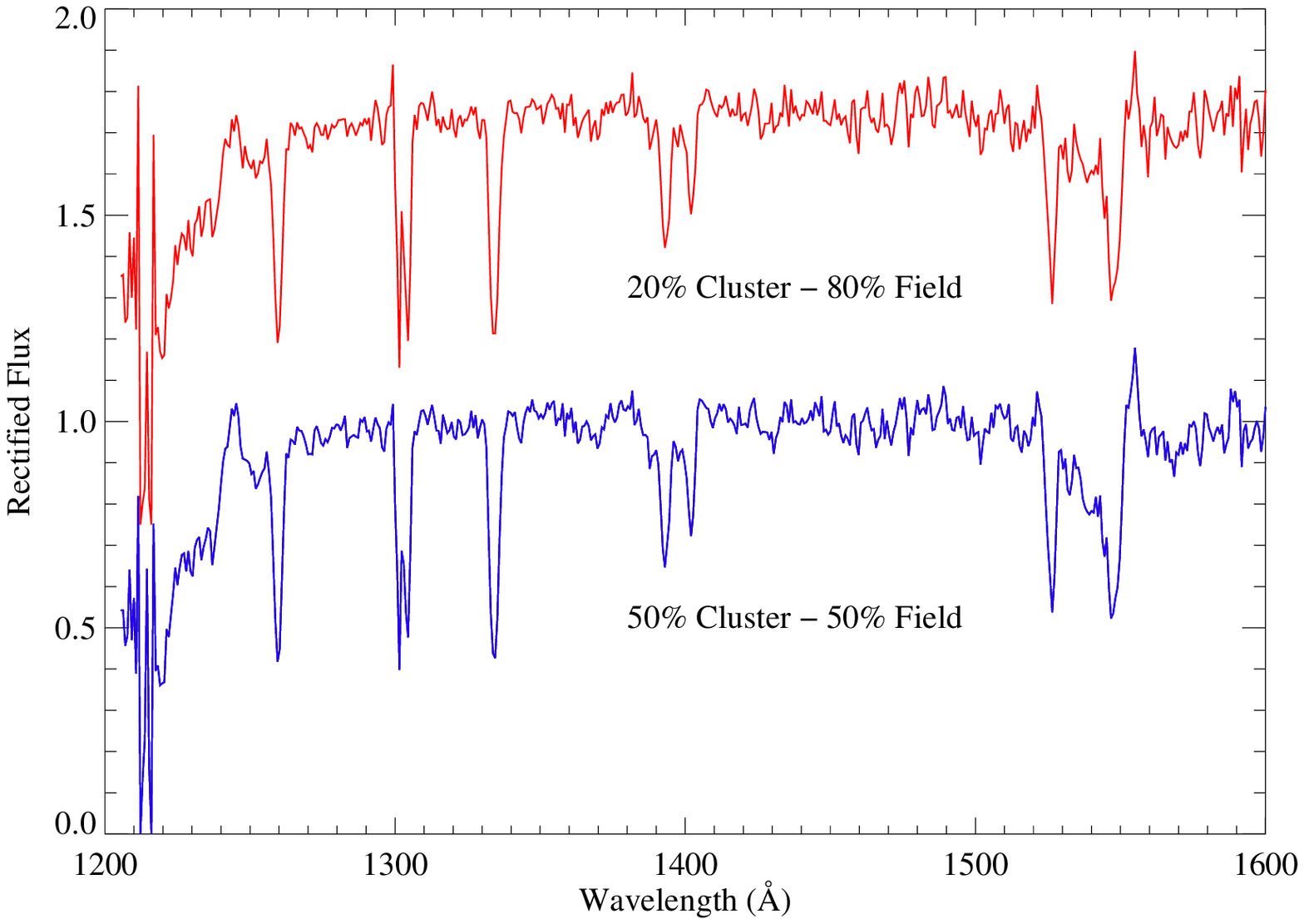}
\caption{We have constructed two spectra to simulate a
spatially integrated spectrum of NGC~5253.  These bracket the observed
values for the fraction of UV light coming from resolved sources: 20\%
clusters -- 80\% field (top) and 50\% clusters -- 50\% field (bottom).  
These template spectra provide a much more accurate benchmark for
comparisons with the spatially integrated spectra of high redshift
galaxies than previously available high resolution starburst spectra,
which are primarily of individual clusters.
\label{fiftyfifty}}
\end{figure}

The differences seen in the spectra of the field and clusters of NGC~5253
have potentially important implications for restframe UV observations of
high redshift galaxies. Because of the small angular sizes of high
redshift objects, the entire star-forming region is typically observed,
with no discrimination between clusters and field. Since the field's UV
light may represent between 50\% and 80\% of the UV output from a
starburst galaxy \citep{Buat_et_al_1994, Meurer_et_al_1995,
Maoz_et_al_1996}, the common tendency to compare spatially integrated
spectra of distant galaxies with the UV spectra of local stellar clusters
neglects the important contribution of the field. To quantify this
statement, we have created two `integrated' spectra, the first combining a
50\% `cluster' spectrum with 50\% `field' spectrum, and the second with a
20\%--80\% cluster--field mix (Figure \ref{fiftyfifty}). The `cluster'
spectrum is the unweighted sum of the eight clusters along our slit prior
to dereddening.  The 50--50 and 20-80 template spectra are available
for download from \url{http://www.stsci.edu/starburst/templ.html}.

The spectral features of the 50--50 `integrated' spectrum
closely resemble those of a continuous starburst with a standard Salpeter
IMF extending up to 100 M$_{\sun}$. The 20--80 integrated spectrum has
weaker wind lines than the 50--50 mix, and may be more appropriate to
populations with declining star formation rates. In both cases the stellar
wind lines, most notably \ion{C}{4}~$\lambda$1550, are substantially
weaker than in a pure young population.  Thus, the generally observed
weakness or even absence of stellar wind lines in high-z galaxies
\citep[e.g.,][]{Lowenthal_et_al_1997} may not be entirely a metallicity
effect, but rather a result of time-extended star-forming episodes.

\begin{figure}
\plotone{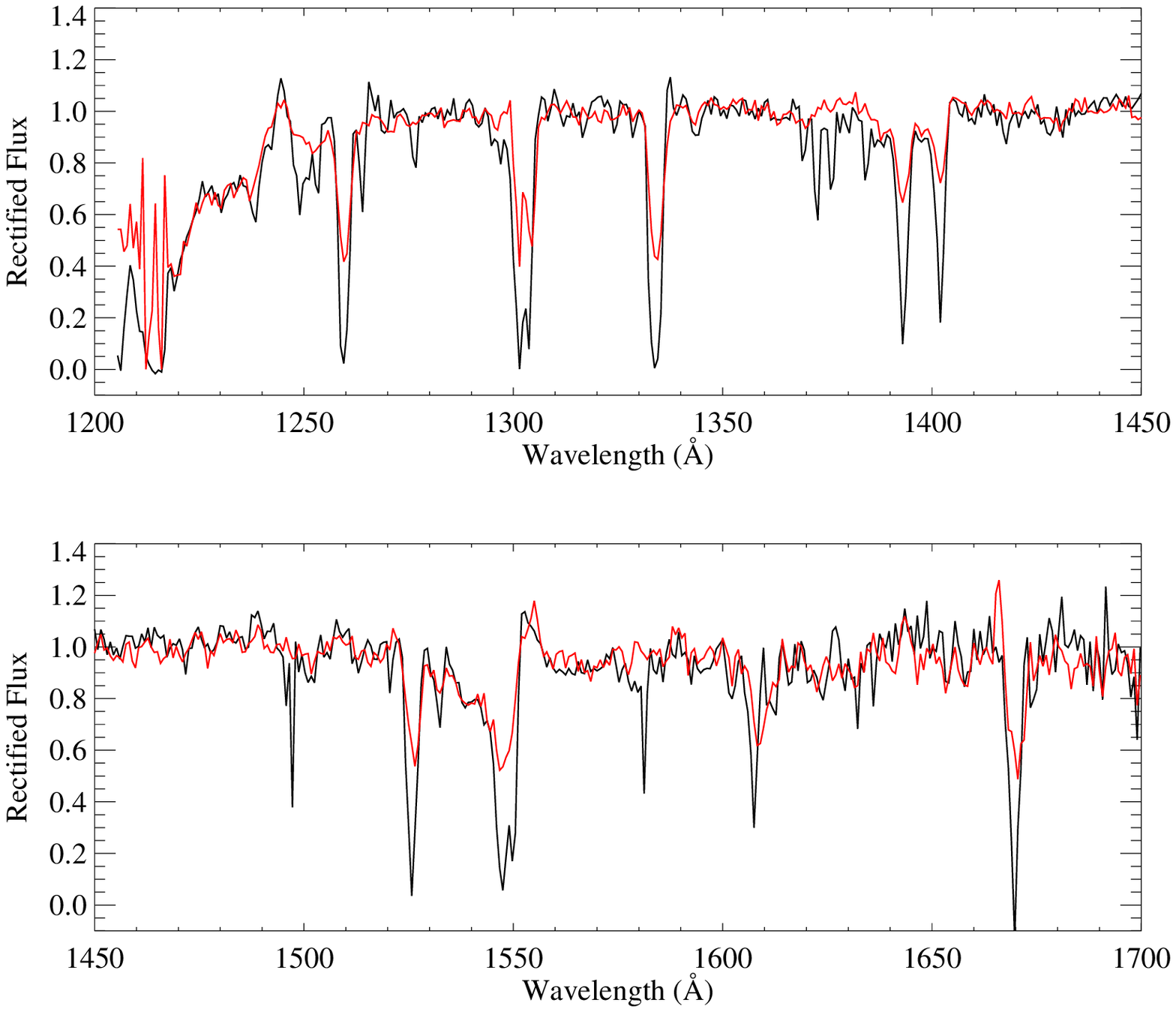}
\caption{The spectrum of MS 1512-cB58, a gravitationally
lensed z = 2.73 starburst, is shown in black \citep{Pettini_et_al_2000}
with the 50\% cluster--50\% field template spectrum of NGC~5253
overplotted in red.  Note the excellent agreement of the continuum and
line profiles of the stellar wind features, \ion{N}{5} $\lambda1240$,
\ion{Si}{4} $\lambda1400$, and \ion{C}{4} $\lambda1550$, with the
exception of the line cores, which are interstellar in origin. All of the
interstellar features of cB58 are much deeper than those of the template
spectrum (and much broader, given that the interstellar lines of NGC~5253
are blended with intervening Milky Way lines); this is probably a result
of the greater covering factor of the gas in the cB58 spectrum, and of the
hydrodynamical consequences of the galaxy's high star formation rate. The
lines present only in the cB58 spectrum near 1375, 1500, and 1580~\AA\ are
intervening absorption features.
\label{cB58}}
\end{figure}

The utility of our template spectra for comparisons with high redshift
galaxies is illustrated in Figure \ref{cB58}, where we compare our 50\%
cluster--50\% field spectrum to the spectrum of the gravitationally lensed
z = 2.73 starburst galaxy, MS 1512-cB58 \citep{Pettini_et_al_2000}.  cB58
appears to be a relatively typical L$^{*}$ Lyman break galaxy, with
Z~$\approx$~1/3~Z$_{\sun}$ \citep{Teplitz_et_al_2000}.  Our empirical
template provides an exceptionally good fit overall to the stellar
features, particularly to the stellar wind lines.  The widths of the
interstellar features appear to be fairly well matched, but it must be
kept in mind that the interstellar features of the NGC~5253 template
spectrum are artificially broadened by blending with intervening Milky Way
absorption features. The intrinsically broader and deeper interstellar
features of cB58 are probably a result of the larger covering factor of
the gas, and the hydrodynamical consequences of the galaxy's high star
formation rate.  Thanks to the exceptional S/N of both cB58 ($\sim40$) and
our template spectrum ($\sim30$), Figure \ref{cB58} provides one of the
best examples to date of the striking similarity between the spectral
morphology of a high redshift star-forming galaxy and a local starburst.

\section{SUMMARY \& CONCLUSIONS}

We have obtained a STIS longslit UV spectrum of the central star forming
region of the dwarf starburst galaxy NGC~5253.  We extracted spectra of 8
clusters and three field regions along the slit, and compared these
spectra to the STARBURST99 stellar evolutionary synthesis models.  The
clusters are well fit by instantaneous burst models with a standard
Salpeter IMF extending from 1--100~M$_{\sun}$.  We derive ages for 
the clusters in the range 1--10 Myr and masses in the 
range 10$^{2}$--10$^{4}$ M$_{\sun}$, although the latter are probably 
underestimated by factors of a few.

The spectral features of the field and the brightest clusters exhibit 
pronounced differences, clear proof that the diffuse UV light is not
simply reflected cluster light.  The main difference is that the field of
NGC~5253 is found to be lacking in the most massive stars, either as a
consequence of aging or because massive stars are not forming in the field
in large numbers.  One possibility is that star formation in starbursts is
bimodal, with compact clusters forming stars with a standard Salpeter IMF
extending up to 100 M$_{\sun}$, and more diffuse star formation occurring
with a massive-star deficient IMF. We suggest an alternate hypothesis
which requires only one `mode' of star formation.  In this scenario,
stellar clusters form continuously and dissolve on $\sim$10~Myr time
scales, dispersing their remaining stars into the field.  We find this
hypothesis to be consistent with the evaporation timescales that we derive for
the majority of our clusters.

Our dissolving clusters scenario suggests that continuous star formation
models with a standard Salpeter IMF in the range 1--100~M$_{\sun}$ should
be appropriate for the integrated spectra of high-z galaxies, which will
contain both young clusters and their dissolved by-products, field stars.  
The commonly observed absence of very broad O-star wind lines in high
redshift galaxies may indicate that we are observing star formation
episodes of sufficient duration to populate the field with substantial
numbers of older stars.  Care must be taken when comparing high resolution
rest-frame UV spectra of distant galaxies to local starburst templates
because the presently available template spectra are primarily of
individual star clusters.  We have constructed a 50\% cluster--50\% field
and a 20\% cluster--80\% field spectrum of NGC~5253 which we present as
new benchmarks for comparison with spectra of high-z galaxies.

\acknowledgements

The authors would like to thank Gerhardt Meurer for insightful and
stimulating discussions which helped shape this paper;
Duilia de Mello for her expert advice about B star lines; and Max
Pettini and Chuck Steidel for contributing the spectrum of MS 1512-cB58.
We also wish to thank our anonymous referee for his or her helpful comments.  
We are grateful for support we received from NASA through Grant No.  
GO-08232.01.97A from the Space Telescope Science Institute which is
operated by the Association of Universities for Research in Astronomy,
Inc. for NASA under contract NAS5-26555.

\clearpage

\clearpage

\begin{deluxetable}{cccccccc}
\tabletypesize{\scriptsize}
\tablecaption{Observed properties of the stellar clusters. 
\label{table_cluster_data}}
\tablewidth{0pt}
\tablehead{
  \colhead{Cluster \tablenotemark{a}} & 
  \colhead{Pixels \tablenotemark{b}} &  
  \colhead{S/N \tablenotemark{c}} &  
  \colhead{R$_{e STIS}$(pc)\tablenotemark{d}} &
  \colhead{R$_{e F170W}$(pc)\tablenotemark{e}} &
  \colhead{F$_{STIS}$\tablenotemark{f}} &
  \colhead{F$_{F170W}$\tablenotemark{g}} &
  \colhead{$\beta$\tablenotemark{h}} 
}

\startdata
2  & 21 & \phn8 & 0.68 & 1.69    & $7.95\times10^{-17}$ & 
  $1.04\times10^{-15}$ & -1.11 \\
3  & 23 & \phn7 & 0.58 & 0.96    & $6.69\times10^{-17}$ & 
  $1.64\times10^{-15}$ & -1.27 \\
5  & 27 & 14    & 2.38 & 4.43    & $2.53\times10^{-16}$ & 
  $1.08\times10^{-15}$ & \phm{-}0.50 \\ 
7  & 29 & 15    & 0.57 & 0.73    & $3.42\times10^{-16}$ & 
  $1.01\times10^{-15}$ & -1.11 \\ 
8  & 13 & \phn8 & 0.17 & \nodata & $7.03\times10^{-17}$ & 
  \nodata & \phm{-}0.13 \\
9  & 23 & \phn9 & 0.59 & \nodata & $9.74\times10^{-17}$ & 
  \nodata & -0.62 \\     
10 & 25 & \phn7 & 0.55 & \nodata & $6.64\times10^{-17}$ & 
  \nodata & -1.14 \\
11 & 27 & \phn6 & 0.70 & \nodata & $3.86\times10^{-17}$ & 
  \nodata & -1.87 \\
\enddata

\tablenotetext{a}{Cluster identification: for clarity we use the naming
convention of Calzetti et al. 1997 for clusters 2, 3, and 5.  We number
the remaining clusters in order of decreasing brightness. The location of
the clusters in the galaxy and relative to the slit is shown in
Figures~\ref{image} and \ref{spatial}.}

\tablenotetext{b}{Number of pixels summed up in the spatial direction for
each cluster. See \S \ref{observations} for details.}

\tablenotetext{c}{Median signal-to-noise of the extracted 1-dimensional
STIS spectrum}

\tablenotetext{d}{Half light radii in parsecs measured by fitting a
gaussian to the cluster spatial profile in the STIS spectrum.  These radii
will underestimate the true value if the slit is not perfectly centered on
the cluster. The value reported for Cluster 8 is for the bright point
source (see \S\ref{clusters}).  1 pc $\approx$ 0\farcs0625 $\approx$ 2.56
pixels.}

\tablenotetext{e}{Half light radii in parsecs measured by fitting a two
dimensional gaussian profile to clusters in the WFPC2 F170W image.  
Clusters 8, 9, 10, \& 11 are too faint to be detected in the image.}

\tablenotetext{f}{Flux in ergs~s$^{-1}$~cm$^{-2}$~\AA$^{-1}$ at 1500~\AA\
measured from the STIS spectrum.  These values have not been corrected for
reddening and slit losses.}

\tablenotetext{g}{Flux in ergs~s$^{-1}$~cm$^{-2}$~\AA$^{-1}$ measured from
photometry of our WFPC2 F170W image. Clusters 8, 9, 10, \& 11 are too
faint to be detected in the image.}

\tablenotetext{h}{Observed power law index of the UV continuum ($F \propto
\lambda^{\beta}$) over the wavelength range 1240 -- 1600~\AA. See
\S\ref{observations} for details.}

\end{deluxetable}

\clearpage

\begin{deluxetable}{cccccccc}
\tabletypesize{\scriptsize}
\tablecaption{Derived properties of the stellar clusters. 
\label{table_cluster_results}}
\tablewidth{0pt}
\tablehead{
  \colhead{Cluster} & 
  \colhead{$\beta_{i}$\tablenotemark{a}} &
  \colhead{E(\bv)\tablenotemark{b}} & 
  \colhead{L$_{STIS}$\tablenotemark{c}} &
  \colhead{L$_{F170W}$\tablenotemark{d}} & 
  \colhead{Age(Myr)\tablenotemark{e}}  & 
  \colhead{Mass(M$_{\sun}$)\tablenotemark{f}} &
  \colhead{Reduced $\chi^2$\tablenotemark{g}} 
}
\startdata
2  & -1.55      & 0.16 & $7.41\times10^{35}$ & $1.17\times10^{37}$ 
& $8 ^{+2.6}_{-0.9}$ & $\phm{>}1\times10^{4}$ & 0.94 \\
3  & -1.71      & 0.14 & $4.87\times10^{35}$ & $1.48\times10^{37}$   
& $3 ^{+0.9}_{-0.9}$  & $\phm{>}4\times10^{3}$ & 1.20 \\
5  & $\phn$0.06 & 0.42 & $2.56\times10^{37}$ & $1.00\times10^{38}$ 
& $2 ^{+0.7}_{-0.8}$ & $\phm{>}4\times10^{4}$ & 1.49 \\
7  & -1.55      & 0.16 & $3.14\times10^{36}$ & $1.11\times10^{37}$ 
& $4 ^{+1.1}_{-0.7}$ & $\phm{>}5\times10^{3}$ & 2.36 \\  
8  & -0.29      & 0.36 & $4.23\times10^{36}$ & \nodata 
& $8 ^{+5.9}_{-1.2}$ & $>5\times10^{3}$ & 1.46 \\
9  & -1.07      & 0.26 & $1.86\times10^{36}$ & \nodata 
& $1 ^{+2.2}$        & $>9\times10^{2}$ & 1.27 \\
10 & -1.48      & 0.18 & $6.82\times10^{35}$ & \nodata 
& $8 ^{+6.5}_{-1.2}$ & $>7\times10^{2}$ & 1.33 \\
11 & -2.31      & 0.05 & $1.17\times10^{35}$ & \nodata 
& $6 ^{+1.2}_{-1.5}$ & $>1\times10^{2}$ & 1.57 \\
\enddata

\tablenotetext{a}{Power law index of the UV continuum ($F \propto
\lambda^{\beta}$) over the wavelength range 1240 -- 1600 \AA\ 
after correcting for a foreground Milky Way reddening of E(\bv) = 0.05.}

\tablenotetext{b}{Reddening derived using the starburst obscuration curve,
assuming an intrinsic slope of $\beta = -2.6$.}

\tablenotetext{c}{Extinction corrected luminosity at 1500 \AA\ in
ergs~s$^{-1}~$\AA$^{-1}$ measured from the STIS spectrum, with no attempt
made at aperture corrections. We use the starburst obscuration curve and
the E(\bv) values quoted here.}

\tablenotetext{d}{Extinction corrected luminosity at 1500 \AA\ in
ergs~s$^{-1}~$\AA$^{-1}$ derived from photometry of a WFPC2 image in the
F170W filter.  The observed slope of the UV continuum is used to correct
the flux given in Table~\ref{table_cluster_data} from the effective
wavelength of the F170W bandpass to 1500 \AA.  Clusters 8, 9, 10, \& 11
are too faint to be measured in the image.}

\tablenotetext{e}{Age derived from the best fitting \emph{STARBURST99}
instantaneous burst model with error bars representing the 90\% confidence
interval. See \S~\ref{modelfitting} for details.}

\tablenotetext{f}{Mass derived by comparing the observed luminosity to
that predicted by \emph{STARBURST99} for a 10$^{6}$ M$_{\sun}$ cluster of
the appropriate age.  For clusters where photometry was unavailable, the
mass is shown as an lower limit, because the aperture corrections are
unknown. Because of uncertainties in the dust geometry and the lower mass
cutoff of the IMF, all of our derived masses may be underestimates by
factors of a few.  See \S~\ref{clusters} for details.}

\tablenotetext{g}{The reduced $\chi^2$ value of the fit of the model to
the cluster spectrum.  See \S~\ref{modelfitting} for details.}

\end{deluxetable}

\clearpage

\begin{deluxetable}{lccccc}
\tabletypesize{\scriptsize}
\tablecaption{Observed properties of the field regions. 
\label{table_field_data}}
\tablewidth{0pt}
\tablehead{
  \colhead{Region \tablenotemark{a}} & 
  \colhead{Pixels \tablenotemark{b}} &  
  \colhead{S/N \tablenotemark{c}} &  
  \colhead{Size(pc$^2$)\tablenotemark{d}} &
  \colhead{F$_{STIS}$\tablenotemark{e}} &
  \colhead{$\beta$\tablenotemark{f}} 
}

\startdata
Field 1     & 127 & 12 & $\phn$80.8  & $2.17\times10^{-16}$ & -1.08 \\
Field 2     & 161 & 16 & 102.4       & $3.47\times10^{-16}$ & -1.20 \\
Field 3     & 167 & 18 & 106.2       & $4.21\times10^{-16}$ & -1.09 \\
Field       & 455 & 27 & 289.4       & $9.80\times10^{-16}$ & -1.12 \\
\enddata

\tablenotetext{a}{Designations given to the three different field regions.  
The row labeled ``Field'' refers to the combined values of the three
regions.  The location of these regions along the slit is shown in Figure
\ref{spatial}.}

\tablenotetext{b}{Number of pixels summed up in the spatial direction for
each field region. See \S \ref{observations} for details.}

\tablenotetext{c}{Median signal-to-noise of the extracted 1-dimensional
STIS spectrum.}

\tablenotetext{d}{Area in parsecs$^{2}$ of each field region.  1\arcsec\
$\approx$ 16 pc.}

\tablenotetext{e}{Flux in ergs~s$^{-1}$~cm$^{-2}$~\AA$^{-1}$ at 1500~\AA\
measured from the STIS spectrum.  These values have not been corrected for
reddening.}

\tablenotetext{f}{Observed power law index of the UV continuum ($F \propto
\lambda^{\beta}$) over the wavelength range 1240 -- 1600~\AA.  See
\S\ref{observations} for details.}

\end{deluxetable} 

\clearpage

\begin{deluxetable}{lcccccccc}
\tabletypesize{\scriptsize}
\tablecaption{Best fitting field models 
\label{table_fieldmodels}}
\tablewidth{0pt}
\tablehead{
  \colhead{Star Formation Law} & 
  \colhead{M$_{up}$\tablenotemark{a}} &
  \colhead{M$_{low}$\tablenotemark{b}} &
  \colhead{$\alpha$\tablenotemark{c}} &
  \colhead{Age(Myr)\tablenotemark{d}}  & 
  \colhead{$\beta_{m}$\tablenotemark{e}} &
  \colhead{E(\bv)\tablenotemark{f}} &
  \colhead{SFR} &
  \colhead{Reduced $\chi^2$\tablenotemark{h}}
}
\startdata
Instantaneous & 100    & 1 & 2.35 & \phn8 & -2.7 & 0.18 & \nodata & 1.8 \\
Continuous    & \phn30 & 1 & 2.35 & 50    & -2.5 & 0.15 & 1.3     & 2.0 \\
Continuous    & 100    & 1 & 3.50 & 50    & -2.1 & 0.08 & 3.0     & 2.2 \\
\enddata

\tablenotetext{a}{Upper mass cut-off of the IMF.} 

\tablenotetext{b}{Lower mass cut-off of the IMF.}

\tablenotetext{c}{Power law slope of the IMF ($\Phi \propto \int
m^{-\alpha}dm$).}

\tablenotetext{d}{Age in Myr of the model used.}

\tablenotetext{e}{Power law index of the UV continuum ($F \propto
\lambda^{\beta}$) of the model over the wavelength range 1240 --
1600~\AA.}

\tablenotetext{f}{Reddening derived using the starburst obscuration curve
from the difference in the model slope, $\beta_{m}$, and the observed
slope of the field after correction for a small amount of foreground
reddening ($\beta = 1.56$). See \S\ref{fieldmodels} for details.}

\tablenotetext{g}{Implied star formation rate in
M$_{\sun}$~yr$^{-1}$~kpc$^{-2}$.} \tablenotetext{h}{The reduced $\chi^2$
value of the fit of the model to the field spectrum.  See
\S~\ref{modelfitting} for details.}

\end{deluxetable}


\clearpage

\begin{thebibliography}{}

\bibitem[Adelberger \& Steidel(2000)]{Adelberger_and_Steidel_2000} 
Adelberger, K.\ L., \& Steidel, C.\ C.\ 2000,  \apj, 544, 218 

\bibitem[Beck et al.(1996)]{Beck_et_al_1996} Beck, S.\ C., Turner, J.\ 
L., Ho, P.\ T.\ P., Lacy, J.\ H., \& Kelly, D.\ M.\ 1996, \apj, 457, 610 

\bibitem[Buat et al.(1994)]{Buat_et_al_1994} Buat, V., Vuillemin, A., 
Burgarella, D., Milliard, B., \& Donas, J. 1994, \aap, 281, 666 

\bibitem[Burstein \& Heiles(1982)]{Burstein_and_Heiles_1982} 
Burstein, D., \& Heiles, C.\ 1982, \aj, 87, 1165 

\bibitem[Caldwell \& Phillips(1989)]{Caldwell_and_Phillips_1989} 
Caldwell, N., \& Phillips, M.\ M.\ 1989, \apj, 338, 789 

\bibitem[Calzetti(1997)]{Calzetti_1997} Calzetti, D.\ 1997, \aj, 113, 
162 

\bibitem[Calzetti et al.(1999)]
{Calzetti_et_al_1999} Calzetti, D., Conselice, C.\ J., Gallagher, J.\ 
S., \& Kinney, A.\ L.\ 1999, \aj, 118, 797 

\bibitem[Calzetti, Kinney \& Storchi-Bergmann(1994)]{Calzetti_et_al_1994} 
Calzetti, D., Kinney, A.\ L., \& Storchi-Bergmann, T.\ 1994, \apj, 429, 582 

\bibitem[Calzetti et al.(1997)]{Calzetti_et_al_1997} Calzetti, D., 
Meurer, G.\ R., Bohlin, R.\ C., Garnett, D.\ R., Kinney, A.\ L., Leitherer, 
C., \& Storchi-Bergmann, T.\ 1997, \aj, 114, 1834 

\bibitem[Calzetti et al.(2000)]{Calzetti_et_al_2000} Calzetti, D., Armus, 
L., Bohlin, R.\ C., Kinney, A.\ L., Koornneef, J., \& Storchi-Bergmann, T.\ 
2000, \apj, 533, 682 

\bibitem[Campbell \& Terlevich(1984)]{Campbell_and_Terlevich_1984} 
Campbell, A.\ W., \& Terlevich, R.\ 1984, \mnras, 211, 15 

\bibitem[Campbell, Terlevich \& Melnick(1986)]{Campbell_et_al_1986} 
Campbell, A., Terlevich, R., \& Melnick, J.\ 1986, \mnras, 223, 811 

\bibitem[Cardelli, Clayton, \& Mathis(1989)]{Cardelli_Clayton_and_Mathis_1989}
Cardelli, J.\ A., Clayton, G.\ C., \& Mathis, J.\ S.\ 1989, \apj, 345, 245 

\bibitem[Garcia-Vargas, Bressan, \& Diaz(1995)]{Garcia-Vargas_et_al_1995} 
Garcia-Vargas, M.\ L., Bressan, A., \& Diaz, A.\ I.\ 1995, \aaps, 112, 13 

\bibitem[Garmany \& Conti(1985)]{Garmany_and_Conti_1985} Garmany, C.\ D., \& 
Conti, P.\ S.\ 1985, \apj, 293, 407 

\bibitem[Gibson et al.(2000)]{Gibson_et_al_2000} Gibson, B.\ K., et al., 
\apj, 529, 723 

\bibitem[Greggio et al.(1998)]{Greggio_et_al_1998} Greggio, L., Tosi, M., 
Clampin, M., de Marchi, G., Leitherer, C., Nota, A.\ \& Sirianni, M.\ 1998, 
\apj, 504, 725 

\bibitem[Heckman et al.(1998)]{Heckman_et_al_1998} Heckman, T.\ M., 
Robert, C., Leitherer, C., Garnett, D.\ R., \& van der Rydt, F.\ 1998, \apj, 
503, 646 

\bibitem[Ho \& Filippenko(1996)]{Ho_and_Filippenko_1996} Ho, L.\ C., \& 
Filippenko, A.\ V.\ 1996, \apj, 472, 600 

\bibitem[Hunter et al.(1996)]{Hunter_et_al_1996} Hunter, D.\ A., O'Neil, 
E.\ J., Lynds, R., Shaya, E.\ J., Groth, E.\ J., \& Holtzman, J\. A.\ 1996, 
\apjl, 459, L27 

\bibitem[Kennicutt et al.(1998)]{Kennicutt_et_al_1998} Kennicutt, R.\ C., 
et al., \apj, 498, 181 

\bibitem[Kim, Morris \& Lee(1999)]{Kim_Morris_and_Lee_1999} Kim, S.\ S., 
Morris, M., \& Lee, H.\ M.\ 1999, \apj, 525, 228 

\bibitem[Kobulnicky et al.(1997)]{Kobulnicky_et_al_1997} Kobulnicky, H.\ A., 
Skillman, E.\ D., Roy, J., Walsh, J.\ R., \& Rosa, M.\ R. 1997, \apj, 477, 679 

\bibitem[Kobulnicky, Kennicutt, \& Pizagno(1999)]
{Kobulnicky_Kennicutt_and_Pizagno_1999} Kobulnicky, H.\ A., 
Kennicutt, R.\ C., \& Pizagno, J.\ L.\ 1999, \apj, 514, 544

\bibitem[Leitherer \& Heckman(1995)]{Leitherer_and_Heckman_1995} 
Leitherer, C., \& Heckman, T.\ M.\ 1995, \apjs, 96, 9 

\bibitem[Leitherer et al.(1999)]{Leitherer_et_al_1999} Leitherer, C., 
et al., 1999, \apjs, 123, 3 

\bibitem[Leitherer et al.(2001)]{Leitherer_et_al_2001} Leitherer, C., 
Le\~{a}o, J.\ R.\ S., Heckman, T.\ M., Lennon, D.\ J., Pettini, M., \&
Robert, C. 2001, ApJ, in press

\bibitem[Lowenthal et al.(1997)]{Lowenthal_et_al_1997} Lowenthal, J.\ D., 
et al., 1997, \apj, 481, 673 

\bibitem[Luhman et al.(2000)]{Luhman_et_al_2000} Luhman, K.\ L., Rieke, 
G.\ H., Young, E.\ T., Cotera, A.\ S., Chen, H., Rieke, M.\ J., Schneider, 
G.\ \& Thompson, R.\ I.\ 2000, \apj, 540, 1016 

\bibitem[Martin \& Kennicutt (1995)]{Martin_and_Kennicutt_1995} Martin, 
C.\ L., \& Kennicutt, R.\ C.\ 1995, \apj, 447, 171 

\bibitem[Massey \& Hunter (1998)]{Massey_and_Hunter_1998} Massey, P., \& 
Hunter, D.\ A.\ 1998, \apj, 493, 180 

\bibitem[Massey et al.(1995)]{Massey_et_al_1995a} Massey, P., Lang, C.\ C., 
Degioia-Eastwood, K., \& Garmany, C.\ D.\ 1995, \apj, 438, 188 

\bibitem[Massey, Johnson, \& Degioia-Eastwood(1995)]{Massey_et_al_1995b} 
Massey, P., Johnson, K.\ E., and Degioia-Eastwood, K.\ 1995, \apj, 454, 151 

\bibitem[Massey(1998)]{Massey_1998} Massey, P. 1998, ASP Conf. 
Ser. 142: The Stellar Initial Mass Function (38th Herstmonceux Conference), 
17 

\bibitem[de Mello, Leitherer, \& Heckman(2000)]{de_Mello_et_al_2000} de 
Mello, D.\ F., Leitherer, C., \& Heckman, T.\ M.\ 2000, \apj, 530, 251 

\bibitem[Meurer, Heckman \& Calzetti(1999)]{Meurer_et_al_1999} Meurer, 
G.\ R., Heckman, T.\ M.\ \& Calzetti, D.\ 1999, \apj, 521, 64 

\bibitem[Meurer et al.(1997)]{Meurer_et_al_1997} Meurer, G.\ R., 
Heckman, T.\ M., Lehnert, M.\ D., Leitherer, C., \& Lowenthal, J.\ 1997, 
\aj, 114, 54 

\bibitem[Meurer et al.(1995)]{Meurer_et_al_1995} Meurer, G.\ R., 
Heckman, T.\ M., Leitherer, C., Kinney, A., Robert, C., \& Garnett, D.\ R.\ 
1995, \aj, 110, 2665 

\bibitem[Maoz et al.(1996)]{Maoz_et_al_1996} Maoz, D., Barth, A.\ J., 
Sternberg, A., Filippenko, A.\ V., Ho, L.\ C., Macchetto, F.\ D., Rix, H. -W., 
\& Schneider, D.\ P.\ 1996, \aj, 111, 2248 

\bibitem[Pettini et al.(2000)]{Pettini_et_al_2000} Pettini, M., Steidel, 
C.\ C., Adelberger, K.\ L., Dickinson, M., \& Giavalisco, M.\ 2000, \apj, 
528, 96 

\bibitem[Puls et al.(1996)]{Puls_et_al_1996} Puls, J., et al. 1996, 
\aap, 305, 171 

\bibitem[Robert, Leitherer, \& Heckman(1993)]
{Robert_Leitherer_and_Heckman_1993} Robert, 
C., Leitherer, C.\, \& Heckman, T. M. 1993, \apj, 418, 749 

\bibitem[Rogstad, Lockart, \& Wright(1974)]{Rogstad_et_al_1974} Rogstad, 
D.\ H., Lockart, I.\ A., and Wright, M.\ C.\ H.\ 1974, \apj, 193, 309 

\bibitem[Sandage \& Brucato(1979)]{Sandage_and_Brucato_1979} Sandage, A., \&
Brucato, R.\ 1979, \aj, 84, 472 

\bibitem[Scalo(1998)]{Scalo_1998} Scalo, J. 1998, ASP Conf. Ser. 
142: The Stellar Initial Mass Function (38th Herstmonceux Conference), 201 

\bibitem[Schaerer et al. (1997)]{Schaerer_et_al_1997} 
Schaerer, D., Contini, T., Kunth, D., \& Meynet, G.\ 1997, \apjl, 481, L75 

\bibitem[Sirianni et al.(2000)]{Sirianni_et_al_2000} Sirianni, M., Nota, 
A., Leitherer, C., De Marchi, G., \& Clampin, M.\ 2000, \apj, 533, 203 

\bibitem[Stasinska \& Leitherer(1996)]{Stasinska_and_Leitherer_1996} 
Stasinska, G., \& Leitherer, C.\ 1996, \apjs, 107, 661 

\bibitem[Steidel et al.(1996)]{Steidel_et_al_1996} Steidel, C.\ C., 
Giavalisco, M., Pettini, M., Dickinson, M., \& Adelberger, K.\ L.\ 1996, 
\apjl, 462, L17 

\bibitem[Steidel et al.(1999)]{Steidel_et_al_1999} Steidel, C.\ C., 
Adelberger, K.\ L., Giavalisco, M., Dickinson, M., \& Pettini, M.\ 1999, 
\apjl, 519, 1 

\bibitem[Strickland \& Stevens(1999)]{Strickland_and_Stevens_1999} 
Strickland, D.\ K., \& Stevens, I.\ R.\ 1999, \mnras, 306, 43 

\bibitem[Teplitz et al.(2000)]{Teplitz_et_al_2000} Teplitz, H.\ I., et al., 
\apjl, 533, L65 

\bibitem[Turner, Beck, \& Ho(2000)]{Turner_Beck_and_Ho_2000} Turner, J.\ L., 
Beck, S.\ C., \& Ho, P.\ T.\ P.\ 2000, \apjl, 532, L109 

\bibitem[Turner, Ho, \& Beck(1998)]{Turner_Ho_and_Beck_1998} Turner, J.\ L., 
Ho, P.\ T.\ P., \& Beck, S.\ C.\ 1998, \aj, 116, 1212 

\bibitem[Walborn et al.(1995)]{Walborn_et_al_1995} Walborn, N.\ R., 
Lennon, D.\ J., Haser, S.\ M., Kudritzki, R., \& Voels, S.\ A.\ 1995, \pasp, 
107, 104 

\bibitem[Walborn et al.(2000)]{Walborn_et_al_2000} Walborn, N.\ R., 
Lennon, D.\ J., Heap, S.\ R., Lindler, D.\ J., Smith, L.\ J., Evans, C.\ 
J., \& Parker, J.\ W.\ 2000, \pasp, 112, 1243 

\bibitem[Walsh \& Roy(1987)]{Walsh_and_Roy_1987} Walsh, J.\ R., \& Roy, 
J.\ 1987, \apjl, 319, L57 

\bibitem[Walsh \& Roy(1989)]{Walsh_and_Roy_1989} Walsh, J.\ R., \& Roy, 
J.\ 1989, \mnras, 239, 297 

\bibitem[Webster \& Smith(1983)]{Webster_and_Smith_1983} Webster, B.\ L., \& 
Smith, M.\ G.\ 1983, \mnras, 204, 743 


\end{thebibliography}
\end{document}